\documentclass[a4paper,11pt]{article}

\usepackage{jheppub} % for details on the use of the package, please
\usepackage[utf8]{inputenc}
\usepackage{hyperref}% add hypertext capabilities
 \usepackage{hyperref}
\usepackage{xcolor,graphicx}% Include figure files
\definecolor{nicegreen}{rgb}{0.1,0.5,0.1}
\definecolor{nicered}{rgb}{0.5,0.1,0.1}
\definecolor{violet}{rgb}{0.7,0.3,0.3}
\hypersetup{colorlinks,citecolor=nicegreen,linkcolor=nicered}

\usepackage{booktabs}
\usepackage{array}
\usepackage{makecell}
\usepackage{graphicx}
\usepackage{natbib}
\usepackage[T1]{fontenc} % if needed
\usepackage{siunitx} % numbers and units
\usepackage{dcolumn}% Align table columns on decimal point
\usepackage{bm}% bold math
\usepackage[normalem]{ulem}%To cross out words with the command \sout{}
\usepackage{braket} % bra-kets for quantum states 
\usepackage{amsmath,mathrsfs}
\usepackage{cancel}%to cross out parts of equations with the command \cancel{}
\usepackage{slashed}
\usepackage{multicol}
\usepackage{multirow}
\usepackage{cleveref}
\usepackage{arydshln}
\usepackage{lscape}
\usepackage{pdflscape}
\usepackage{easyReview}

\newcommand{\cL}{\mathscr{L}}
\newcommand{\cO}{\mathcal{O}}

\renewcommand{\L}{\mathscr{L}}

\newcommand{\Amp}{\mathcal{A}}

\def\INFNPisa{INFN, Sezione di Pisa, Largo Bruno Pontecorvo 3, I-56127 Pisa, Italy }
\def\Unipi{Dipartimento di Fisica, Universit\`a di Pisa, Largo Bruno Pontecorvo 3, I-56127 Pisa, Italy }
\def\RomaTre{Università degli Studi and INFN Roma Tre, Via della Vasca Navale 84, I-00146, Rome, Italy }
\def\CERN{Theoretical Physics Department, CERN, 1211 Geneva 23, Switzerland}

\title{Multiboson production at multi-TeV lepton colliders as complementary probes to rare Higgs decays}
\author[1]{Dario Buttazzo,}
\author[1,2]{Keshav Chaudhary,}
\author[3,4]{Roberto Franceschini,}
\author[1]{Arman Korajac}

\affiliation[1]{\INFNPisa}
\affiliation[2]{\Unipi}
\affiliation[3]{\RomaTre}
\affiliation[4]{\CERN}

\emailAdd{dario.buttazzo@pi.infn.it}
\emailAdd{keshav.chaudhary@student.unipi.it}
\emailAdd{roberto.franceschini@uniroma3.it}
\emailAdd{arman.korajac@pi.infn.it}
\begin{document}

\abstract{
We study the potential of very high-energy lepton colliders to probe effective operators affecting the loop-induced Higgs boson decays $h\to \gamma\gamma$ and $h\to Z\gamma$, providing constraints complementary to Higgs coupling measurements at the Large~Hadron~Collider. In particular, we consider processes with two and three bosons in the final state and analyze their individual and combined impact on the effective field theory parameter space. We demonstrate that these processes at a multi-TeV lepton collider can improve the bounds from rare Higgs decays and electroweak precision observables, surpassing the projections for High-Luminosity LHC and FCC-ee. This is especially true for triboson processes, whose sensitivity is enhanced by the growth with energy of the new physics cross-sections, making them the dominant constraint at very high energies, potentially competitive with FCC-hh.}

{\tiny{\hfill CERN-TH-2026-257}}

\maketitle
\newpage
%%%%%%%%%%%%%%%%%%%%%%%%%%%%%%%%%%%%%%%%%%%%%%%%%%%%%%%%%%%%%%%%%%%%%%%%%%%%%%%%%%%%%%%%%%%%%%%%%%%%

%%%%%%%%%%%%%%%%%%%%%%%%%%%%%%%%%%%%%%%%%%%%%%%%%%%%%%%%%%%%%%%%%%%%%%%%%%%%%%%%%%%%%%%%%%%%%%%%%%%%
\section{Introduction}
\label{sec:intro}
%%%%%%%%%%%%%%%%%%%%%%%%%%%%%%%%%%%%%%%%%%%%%%%%%%%%%%%%%%%%%%%%%%%%%%%%%%%%%%%%%%%%%%%%%%%%%%%%%%%%

The LHC has established the Standard Model (SM) as a theory suitable to describe all available data in the energy domain up to a few TeV probed in its $pp$ collisions. Nevertheless, there are reasons to expect New Physics (NP) beyond the Standard Model appearing at not too much higher energies.
The main issue has to do with the origin of the weak scale, which in quantum field theory appears to be extremely sensitive to the presence of NP at any scale, provided it couples to the SM. If new physics appears at much higher energies, this leads to the puzzle of why the weak scale remains so much smaller instead of being driven towards the   high scale. The absence of NP up to the few-TeV energies probed by the LHC thus gives rise to a tension with the observed value of the weak scale.

In this picture the Higgs boson is our keyhole from which we can observe the physics of a higher scale, that somehow fails to cause a large shift in the weak scale, despite its ability to potentially do so.
The Higgs boson thus deserves the highest degree of attention, and its properties must be studied in the finest detail and from all angles, as to identify any deviation from the SM. This can be done in two complementary ways: collecting large amounts of Higgs bosons to study its properties with increased precision, and exploiting high energies where NP effects become larger.
Rare processes, such as its loop-induced decay modes, are particularly promising in this respect, as NP could contribute with relatively much larger new effects that are easier to spot experimentally. 
      
In the cases where the NP interactions arise at a higher scale or with a smaller coupling, and are particularly elusive in other observables,
the production of massive amounts of Higgs bosons and their decay into rare final states, notably $h\to \gamma\gamma$ and $h\to Z\gamma$, is today the main way forward to test NP. These decays will benefit significantly from the increase of luminosity brought by the High Luminosity LHC (HL-LHC). At a quantitative level our knowledge of Higgs boson interactions at the end of HL-LHC will be roughly as follows. The decay rates of $h \to WW$ and $ h \to ZZ$ will be measured down to 1\% precision. The $h \to \gamma\gamma$ decays will also be known down to 1.5\%, while $h\to\gamma Z$ will be known with a lesser precision of around 6\%. 
At FCC-ee and other leptonic Higgs factories there is going to be progress on the most copious decay rates, e.g. $h \to b \bar b$, but the rate is going to be too small to probe rare decays. As a consequence, the advance of our knowledge of these channels will be deferred to future $pp$ colliders. At FCC-hh, thanks to its huge Higgs production rates, precisions of around 0.3\% and 0.7\% can be achieved for the two decays $h\to \gamma\gamma$ and $h\to Z\gamma$, respectively~\cite{deBlas:2025gyz}.
A relevant question then is whether complementary high-energy probes can be conceived, that could compete with the precision that will be achieved by HL-LHC in these rare decays.
      
\medskip 

\begin{figure}
  \centering
  \raisebox{-0.5\height}{%
    \includegraphics[width=0.48\linewidth]
      {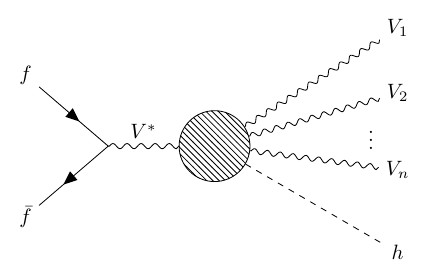}%
  }
  \hfill
  \raisebox{-0.5\height}{%
    \includegraphics[width=0.48\linewidth]
      {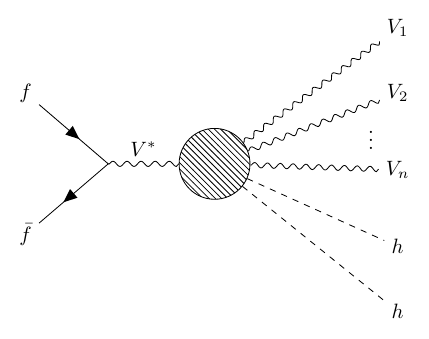}%
  }
  \caption{Sketch of the processes involving the energy-sensitive EFT interactions $\cO_{HWB}, \cO_{HB}, \cO_{HW}$.}
  \label{fig:sketch-processes}
\end{figure}

In this work, we study high-energy effects induced by the same interactions that affect the rare Higgs decays, in order to probe the same physics in a complementary way. We shall adopt the framework of the SM Effective Field Theory (SMEFT), where the decays $h\to \gamma\gamma$ and $h\to Z\gamma$ are mediated by the operators $\cO_{HWB}, \cO_{HB}, \cO_{HW}$ defined in \cite{Grzadkowski:2010es} and recalled below in \cref{eq:opHVVSMEFT}. The same interactions can also mediate processes involving more legs, specifically $2 \to 2$ and $2 \to 3$ scatterings. We will study for the first time the production of multiple Higgs and vector bosons induced by these operators, and show that the rates of these scatterings put some of them at a competitive level of constraining power for the EFT interactions.
In particular, we shall see that the scattering processes are in general dependent on different combinations of the Wilson coefficients than those that enter in the rare Higgs boson decays, and as such provide complementary information. This can lead to substantial improvements in the global studies of Higgs SMEFT interactions. A sketch of possible scattering processes that directly probe those interactions is shown in the Feynman diagrams of \Cref{fig:sketch-processes}.

We take a high-energy lepton collider, of which the prime example is a 10~TeV muon collider, as the main target for the study of these multi-body scatterings. Due to the elementary nature of the leptons, most of the beam energy is available for the hard scatterings, allowing high-momentum transfer vertices involving gauge bosons.
This significantly enhances the SMEFT interactions under consideration, which are sensitive to the transferred momentum, as opposed to the energy-insensitive renormalizable couplings of the SM.

This principle of ``energy helps accuracy'' has been exploited already for the muon collider in~\cite{Buttazzo:2020uzc}, and earlier for several LHC applications for instance in~\cite{Franceschini:2017xkh,Panico:2017frx,Farina:2016rws}. These earlier works showed that the study of fixed-energy properties, such as decay rates, can be complemented, in some cases surpassed, by the study of high-energy processes. Here we extend the analysis to the three operators entering the rare Higgs decays, whose different high-energy behaviour will lead us to go beyond diboson final states. The exploration of these scatterings at hadron colliders is also possible, but we leave it for future work. Similarly, we shall propose some indicative results for lepton collisions at energy different from 10~TeV, but we shall conduct a detailed study only for 10~TeV.

%%%%%%%%%%%%%%%%%%%%%%%%%%%%%%%%%%%%%%%%%%%%%%%%%%%%%%%%%%%%%%%%%%%%%%%%%%%%%%%%%%%%%%%%%%%%%%%%%%%%
%%%%%%%%%%%%%%%%%%%%%%%%%%%%%%%%%%%%%%%%%%%%%%%%%%%%%%%%%%%%%%%%%%%%%%%%%%%%%%%%%%%%%%%%%%%%%%%%%%%%

This paper is organized as follows. In \Cref{sec:EFT} we introduce the EFT couplings that we intend to study. In \Cref{sec:precision} we draw the landscape of precision studies on radiative Higgs couplings at the LHC and future colliders, and we outline the problem of flat directions in the SMEFT parameter space. In \Cref{sec:diboson} we derive the constraints from diboson production at high-energy lepton colliders, while in \Cref{sec:triboson} we study triboson production and summarize our combined results. Finally we give our conclusions in \Cref{sec:conclusions}.

%%%%%%%%%%%%%%%%%%%%%%%%%%%%%%%%%%%%%%%%%%%%%%%%%%%%%%%%%%%%%%%%%%%%%%%%%%%%%%%%%%%%%%%%%%%%%%%%%%%%
\section{Effective Field Theory description}
\label{sec:EFT}
%%%%%%%%%%%%%%%%%%%%%%%%%%%%%%%%%%%%%%%%%%%%%%%%%%%%%%%%%%%%%%%%%%%%%%%%%%%%%%%%%%%%%%%%%%%%%%%%%%%%

Effects of heavy states from a Standard Model (SM) ultraviolet extension can be encapsulated in higher-dimensional ($d > 4$) local effective operators in the Standard Model Effective Field Theory (SMEFT) framework. 
Restricting ourselves to the lowest-dimensional gauge-invariant operators that affect the loop-induced Higgs couplings to the electroweak gauge bosons, a suitable description of these phenomena at high-energy colliders is given by the SM Lagrangian complemented with the following dimension-six operators given in the Warsaw basis \cite{Grzadkowski:2010es}:
\begin{align}\label{eq:opHVVSMEFT}
\cO_{HWB} &= H^\dagger \sigma^{a} H W_{\mu\nu}^a B^{\mu \nu}, &
\cO_{HB} &= |H|^2 B_{\mu\nu}B^{\mu\nu}, &
\cO_{HW} &= |H|^2 W^a_{\mu\nu}W^{a \,\mu\nu},
\end{align}
where $a = 1,2,3$ is the $SU(2)_L$ index, $\sigma^a$ are the Pauli matrices, $H$ is the Higgs doublet field, and $W_{\mu\nu}^a, B_{\mu \nu}$ are the $SU(2)_L$ and $U(1)_Y$ field strength tensors, respectively. An analogous set of CP-violating operators $\widetilde\cO_{HB} = |H|^2 B_{\mu\nu}B_{\rho\sigma}\epsilon^{\mu\nu\rho\sigma}$ (and similar for $W$) can be defined, but we will neglect them in the following, assuming CP-invariance of new physics.
These operators enter the interaction Lagrangian weighted by their Wilson coefficients $c_i$,
\begin{align}\label{eq:lagrHVVSMEFT}
\cL_{\mathrm{NP}} \supset \frac{c_{HWB}}{\Lambda^2}\cO_{HWB}  + \frac{c_{HB}}{\Lambda^2}\cO_{HB} + \frac{c_{HW}}{\Lambda^2}\cO_{HW},
\end{align}
where $\Lambda$ is the typical energy scale of NP in units of mass/coupling.

The operators in~\cref{eq:lagrHVVSMEFT} will contribute to different processes, such as the rare Higgs decays $h \to \gamma \gamma $ and $h \to Z \gamma$, along with other high- and low-energy observables. There are several reasons that make a study of these interactions interesting.
As these decays arise at loop-level and are suppressed in the SM, they have a chance to be sensitive to higher new physics scales $\Lambda$, as the NP effects may be easier to detect compared to other processes where SM contributions arise at tree-level. This fact is especially true if the operators in \cref{eq:lagrHVVSMEFT} are generated from tree-level interactions in the UV theory. 

Independently  of the UV origin of the interactions in \cref{eq:lagrHVVSMEFT}, it is important to realize that these operators also affect other observables than the Higgs rare decay rates, such as Higgs production rates, electroweak precision observables, and high-energy scattering cross-sections.
In the following we will focus on the scattering rates for leptons into multiple gauge and Higgs boson final states as sketched in \Cref{fig:sketch-processes}. Our goal is to illustrate the complementarity and interplay between the Higgs pole observables and the high-energy ones.

We remark that in a fully general SMEFT context other operators can be present, but in concrete UV realizations only (small) subsets are expected to have sizeable Wilson coefficients. 
In particular, operators of the type $|H|^6$ and $|H|^4 D^2$ exert an indirect impact on the processes we study by modifying the Higgs vacuum expectation value and shifting the Higgs field. We do not consider the effects of these operators in this work, as we will focus on high-energy effects. In any case they have been studied in~\cite{Brivio:2017bnu, Brivio:2017vri} and can be constrained well from other observables.

The simultaneous presence of operators of different type in general depends on the type of UV theory under consideration. It is worth asking in which scenarios the operators in \cref{eq:lagrHVVSMEFT} can be generated without sizable contributions to the other dimension-6 SMEFT operators that we neglect.
As a reference example for a microscopic scenario of this type, one can consider an extension of the SM with new electroweak matter of mass $M$, subject to some kind of matter parity that forbids it to couple to the SM linearly. This would for instance be the case for a dark matter candidate particle. If this new electroweak matter is also coupled to the Higgs boson with coupling $\lambda$, then one in general expects the new matter running in box-type loops to give rise to~\cref{eq:lagrHVVSMEFT} with
\begin{align}
     c_{i} &\approx \frac{g_{\rm SM}^2\lambda^2}{16 \pi^2}\,, & \Lambda &\approx M\,,
     \label{eq:power-counting}
\end{align}
where $g_{\rm SM} = g_1, g_2$ is the relevant SM gauge coupling.
An explicit class of models of this type is the one of supersymmetric extensions of the SM, in which there are contributions to \cref{eq:lagrHVVSMEFT} from loops of SUSY partners of leptons, for which $\lambda$ would be replaced by a combination of A-terms and Yukawa couplings. The analog of this scenario with squarks in place of sleptons has been studied in Refs.~\cite{Drozd:2015kva,Kraml:2025fpv}. Other operators would be generated as well in such scenario, hence the study of the effects tied to \cref{eq:lagrHVVSMEFT} should be thought of as a part of more global effort. Notice that in all these examples the operators in \cref{eq:opHVVSMEFT} are generated at one loop, so the constraints that we derive in the next Sections must be weakened by a factor of $4\pi$.

Obviously, the validity of the EFT expansions is more at risk when   new physics effects also arise at loop level. We will discuss the issue of EFT validity in more detail in \Cref{sec:EFTvalidity}, keeping in mind the possible tree- and loop-level origin of the new physics hidden inside the Wilson coefficients. More exotic examples of models where the operators in \cref{eq:opHVVSMEFT} are generated at tree-level are possible, and have e.g.\ been presented in~\cite{Craig:2019wmo,Barducci:2025twe}.

%%%%%%%%%%%%%%%%%%%%%%%%%%%%%%%%%%%%%%%%%%%%%%%%%%%%%%%%%%%%%%%%%%%%%%%%%%%%%%%%%%%%%%%%%%%%%%%%%%%%
\section{Precision constraints}
\label{sec:precision}
%%%%%%%%%%%%%%%%%%%%%%%%%%%%%%%%%%%%%%%%%%%%%%%%%%%%%%%%%%%%%%%%%%%%%%%%%%%%%%%%%%%%%%%%%%%%%%%%%%%%

In this section we discuss the current and projected constraints from precision Higgs and electroweak measurements. For technical clarity we specify that throughout the work the $\{m_W, m_Z, G_F\}$ \cite{Brivio:2017bnu, Brivio:2020onw} input scheme is used, where $m_W$ is the $W$-boson mass, $m_Z$ is the $Z$-boson mass, and $G_F$ is Fermi's constant.

%%%%%%%%%%%%%%%%%%%%%%%%%%%%%%%%%%%%%%%%%%%%%%%%%%%%%%%%%%%%%%%%%%%%%%%%%%%%%%%%%%%%%%%%%%%%%%%%%%%%
\subsection{Higgs decays}
\label{sec:higgs}

We start by discussing the decay $h \to \gamma \gamma$, which proceeds at loop-level in the SM, where it is mediated by $W$ boson and top quark loops.
The decay width at leading order in SM is given by \cite{Djouadi:2005gi}
\begin{equation}\label{eq:higgsGammaGamma:SMxsec}
\Gamma^{\rm SM} (h \to \gamma \gamma) = 
\frac{\sqrt{2}G_F \alpha_{\rm em}^2 m_h^3}{16\pi^3}|\mathcal{I^\gamma}|^2\,,
\end{equation}
where $\alpha_{\rm em} = e^2/4\pi$ is the fine structure constant, $m_h$ is the Higgs boson mass, and the approximate value of the SM loop function is $\mathcal{I}^\gamma \simeq -1.65$ \cite{Manohar:2006gz}. 
The modified decay width with respect to the SM value is \cite{Manohar:2006gz, Alonso:2013hga, Brivio:2017vri}
\begin{equation}
    |\kappa_{\gamma}|^2=\frac{\Gamma(h \to \gamma \gamma)}{\Gamma^{\rm SM}(h \to \gamma \gamma)} = \Big| 1+ \frac{8 \pi^2}{\mathcal{I}^\gamma}\frac{v^2}{\Lambda^2} \mathcal{C}_{\gamma\gamma} \Big|^2 + \Big| \frac{8 \pi^2}{\mathcal{I}^\gamma} \frac{v^2}{\Lambda^2} \tilde{\mathcal{C}}_{\gamma\gamma} \Big|^2\,,
\end{equation}
where
\begin{align}
\mathcal{C}_{\gamma\gamma}  = \frac{1}{g_2^2} c_{HW} + \frac{1}{g_1^2} c_{HB} - \frac{1}{g_1 g_2} c_{HWB} \,,
\end{align}
with $g_1$ the $U(1)_Y$ coupling, $g_2$ the $SU(2)_L$ coupling, and a similar expression holds for $\tilde{\mathcal{C}}_{\gamma\gamma}$ in terms of the $\tilde{c}$ Wilson coefficients, which we however omit for the remainder assuming invariance under CP.

The Higgs decay $h\to Z\gamma$ is primarily dominated by the $W$ boson loop, with only a minor contribution coming from the top quark loop. In the SM, the decay width is given by the following expression~\cite{Bergstrom:1985hp}:
\begin{equation}
\Gamma^{\text{SM}}(h \to Z\gamma) = \frac{\sqrt{2}G_F \alpha_{\rm em}^2 m_h^3}{8\pi^3} \left( 1 - \frac{m_Z^2}{m_h^2} \right)^3 \left| \mathcal{I}^{Z\gamma} \right|^2,
\end{equation}
where $\mathcal{I}^{Z\gamma} \approx -2.87$ \cite{Manohar:2006gz, Bergstrom:1985hp}. The decay rate in the SMEFT has the form~\cite{Alonso:2013hga, Manohar:2006gz}
\begin{equation}
|\kappa_{Z\gamma}|^2 = \frac{\Gamma(h \to Z\gamma)}{\Gamma^{\text{SM}}(h \to Z\gamma)} = \left| 1 + \frac{8\pi^2}{\mathcal{I}^{Z\gamma}}\frac{v^2}{\Lambda^2} \mathcal{C}_{Z\gamma} \right|^2 + \left| \frac{8\pi^2}{\mathcal{I}^{Z\gamma}}\frac{v^2}{\Lambda^2} \tilde{\mathcal{C}}_{Z\gamma} \right|^2,
\end{equation}
where
\begin{align}
\mathcal{C}_{Z\gamma} &= \frac{1}{g_1 g_2} c_{HW} - \frac{1}{g_1 g_2} c_{HB} - \left( \frac{1}{2 g_1^2} - \frac{1}{2 g_2^2} \right) c_{HWB},
\end{align}
and similarly for $\tilde{\mathcal{C}}_{Z\gamma}$ (which we shall neglect in the following).

The current bounds from ATLAS and CMS on the Higgs modified couplings $\kappa_\gamma$ and $\kappa_{Z\gamma}$ in the so-called $\kappa$ framework \cite{ParticleDataGroup:2024cfk, ATLAS:2022tnm, CMS:2022dwd} are given in \Cref{tab:bounds} together with a single-operator translation in terms of the Wilson coefficients of our interest.
In the following, we will use the following constraints on the modified couplings,
\begin{align}
    \kappa_\gamma &= 1.04\pm 0.05\,,\\
    \kappa_{Z\gamma} &= 1.50^{+0.23}_{-0.25}\,,
\end{align}
which are based on a combined fit \cite{Barlow:2024iwp} to ATLAS and CMS results in the $\kappa$ framework. This uncertainty on $\kappa_\gamma$ is projected to be reduced to 1.5\% during the high-luminosity LHC phase~\cite{deBlas:2025gyz,Altmann:2025feg}.
The larger uncertainty for $\kappa_{Z\gamma}$ shows that the bound from the diphoton channel will be more stringent. Future projections~\cite{deBlas:2025gyz,Altmann:2025feg} suggest the uncertainty on $\kappa_{Z\gamma}$ to be reduced to about 6\% at the high-luminosity LHC. Owing to the correlation between the two decay channels, a joint fit may improve the sensitivity to the operator parameter space. However, an additional experimental input is required for the parameter space to be constrained.

ATLAS has also presented observed and expected one-dimensional bounds on the SMEFT Wilson coefficients in \cite{ATLAS:2022tnm}, obtained from a combined fit of the modified Higgs signal strength $\mu_{\gamma \gamma}$ across different template cross section regions. Our results (shown in the lower part of \Cref{tab:bounds}), derived from a simple combined fit from the bounds on $\kappa_\gamma$ and $\kappa_{Z\gamma}$, differ by at most $\sim 20\%$ from the full-fledged analysis results of ATLAS. This primarily reflects the fact that the constraints are largely driven by the measurement of the $h \to \gamma\gamma$ branching ratio, thus justifying our simplification. We find that the main difference lies in the central values, while the overall sensitivities remain comparable.
Overall we consider such agreement satisfactory for our purpose of casting sensitivity projections, hence we adopt the simplified fit based on the inclusive $\kappa_\gamma$. This has two practical consequences.
First, the bounds on the SMEFT operators derived in \cite{ATLAS:2022tnm} are provided in a single-operator framework. Using the simplified fit based on $\kappa_\gamma$ we can obtain   multi-dimentional constraints, as needed for our analysis. Second, this approach facilitates a consistent comparison between current and projected bounds on the Wilson coefficients, which are presented from typically less refined analyses than the current one.

%%%%%%%%%%%%%%%%%%%%%%%%%%%%%%%%%%%%%%%%%%%%%%%%%%%%%%%%%%%%%%%%%%%%%%%%%%%%%%%%%%%%%%%%%%%%%%%%%%%%
\begin{table}
\renewcommand{\arraystretch}{1.35}
    \centering
    \begin{tabular}{|c|c|c|c|}
    \hline\hline
        & ATLAS $\kappa$ & CMS $\kappa$ & ATLAS SMEFT \\
         \hline
        $\kappa_\gamma$ & $1.01\pm0.06$ & $1.10 \pm 0.08$ & \multirow{2}{*}{\centering  --}\\
        $\kappa_{Z\gamma}$ & $1.38^{+0.31}_{-0.37}$ & $1.65^{+0.34}_{-0.37}$ & \\
        \hline
        $c_{HW}$ &$-0.003\pm 0.017$ & $-0.018 \pm 0.023$ & $-0.0035^{+0.014}_{-0.016}$\\
        $c_{HB}$ &$-0.0003 \pm 0.0052$ & $-0.004 \pm 0.007$ & $-0.0011^{+0.0044}_{-0.0050}$\\
        $c_{HWB}$ & $0.001 \pm 0.009$& $0.008 \pm 0.013$ & $0.0020^{+0.0090}_{-0.0079}$\\
    \hline\hline
    \end{tabular}
    \caption{\textbf{Upper part:} Bounds on the modified couplings $\kappa$ given by ATLAS \cite{ATLAS:2022vkf} and by CMS \cite{CMS:2022dwd}. \textbf{Lower part:} One-dimensional bounds on the Wilson coefficients from the modified couplings $\kappa$ by ATLAS (first column), by CMS (second column), as well as the bounds derived by the full SMEFT analysis by ATLAS (last column) \cite{ATLAS:2022tnm}. The scale $\Lambda$ is set to 1 TeV.}
    \label{tab:bounds}
\end{table}
%%%%%%%%%%%%%%%%%%%%%%%%%%%%%%%%%%%%%%%%%%%%%%%%%%%%%%%%%%%%%%%%%%%%%%%%%%%%%%%%%%%%%%%%%%%%%%%%%%%%

%%%%%%%%%%%%%%%%%%%%%%%%%%%%%%%%%%%%%%%%%%%%%%%%%%%%%%%%%%%%%%%%%%%%%%%%%%%%%%%%%%%%%%%%%%%%%%%%%%%%
\subsection{Electroweak precision measurements}
\label{sec:ewpt}

The operators $\cO_{HB}$ and $\cO_{HW}$ modify the kinetic terms of the gauge fields, which can be reabsorbed by a proper redefinition of the fields and the gauge couplings \cite{Barbieri:2004qk,Brivio:2020onw}. 
Therefore, they do not enter into radiative corrections of the gauge boson propagators, nor they modify the gauge boson couplings to fermions. Ultimately, this leads to the conclusion that they do not affect electroweak precision observables (EWPO) \cite{Barbieri:2004qk,Ciuchini:2013pca}. In contrast, $\cO_{HWB}$ generates a non-diagonal kinetic term proportional to $B^{\mu \nu}W^3_{\mu \nu}$, requiring a field rotation to diagonalize the gauge sector. As a result, electroweak input parameters acquire corrections proportional to $c_{HWB}$, rendering the EWPO fit dependent on it.
We discuss below in detail the dependence of these observables on the SMEFT operators.

In particular, $\cO_{HWB}$ induces modifications of the couplings of gauge bosons to fermions. The modified $Z$ couplings directly enter the LEP $Z$-pole measurements. In the $\{m_W, m_Z, G_F\}$ scheme, the photon coupling is also modified, corresponding to a shift in the electromagnetic coupling $\alpha_{\rm em}$. In addition, the triple gauge couplings $WWZ$ and $WW\gamma$ are modified by $\cO_{HWB}$~\cite{Brivio:2020onw}.\footnote{The $f\bar{f}\gamma$ and $WW\gamma$ couplings modifications compensate each other to preserve electromagnetic gauge-invariance in physical observables.}

Combining the $e^+e^-\to \bar{f}f$ and $e^+e^-\to W^+W^-$ observables measured at LEP~\cite{ALEPH:2005ab}, 
the following bound on the operator $c_{HWB}$ is found at 95 \% CL in the $m_W$ input scheme~\cite{Brivio:2017bnu}:
\begin{equation}
    c_{HWB} \in [-0.15, 0.06] \times 10^{-2} (\Lambda^2/v^2)\,,
\end{equation}
derived with a one-parameter fit and zero theoretical uncertainty due to neglected effects of higher-order operators in the analysis. 
For later convenience we give the bound for the scale $\Lambda = 20 $ TeV and we find
\begin{equation}\label{eq:EWPOcurrent}
    c_{HWB} \in [-10.0, 4.0]  
\end{equation}
at 95\% CL. Using preliminary FCC-ee pseudodata and HL-LHC projections, the following projection of bounds can be derived \cite{Celada:2024mcf}:
\begin{equation}
    \left|\frac{c_{HWB}}{\Lambda^2}\right| < 1.2 \times 10^{-3}  \,\mathrm{TeV}^{-2}\,,
\end{equation}
and for $\Lambda = 20 $ TeV we find
\begin{equation}\label{eq:EWPOprojected}
    |c_{HWB}| < 0.48 \,,
\end{equation}
which suggest an improvement by an order of magnitude in comparison to current limits.

%%%%%%%%%%%%%%%%%%%%%%%%%%%%%%%%%%%%%%%%%%%%%%%%%%%%%%%%%%%%%%%%%%%%%%%%%%%%%%%%%%%%%%%%%%%%%%%%%%%%
\subsection{Bounds on NP from global fits}
\label{sec:global-fits}

The above effects above in Higgs decays and traditional low-energy EWPT are the main routes to test the operators \cref{eq:opHVVSMEFT} presently envisaged at LHC and future colliders. Depending on the UV origin of new physics these operators may or may not be studied in isolation with respect to other EFT interactions. To bridge our ignorance about the UV origin of new physics it has become customary to perform global analyses of SMEFT effects in which many EFT interactions are considered simultaneously, and a large amount of data from independent observables is used to put constraints on all the EFT interactions at once. These exercises may not necessarily have an easy to guess UV backing, but serve the purpose to ensure that present and future experimental efforts have the constraining power to study a very broad class of UV models through the EFT approach. In particular, global constraints exercises help to highlight cases in which effects of new physics can partly escape detection due to cancellations of their effects in some observables. When such cancellations happen, there is a so-called ``flat-direction'' in the space of couplings in which there is little constraining power on new physics from the data under study. In such cases, the need arises for new pieces of data to be added to the analysis. With our study, among other things, we aim at identifying such flat directions within the set of operators of \cref{eq:opHVVSMEFT} and provide possible ways to eliminate them.
Before discussing the details, we summarize here the landscape of probes presently considered in SMEFT global fits for the LHC and future colliders.

Global analyses of Higgs couplings probes at the LHC and future lepton colliders have been performed in  \cite{DeBlas:2019qco,deBlas:2022ofj}, combining Higgs and electroweak precision observables in a global fit, from which the sensitivity on modified couplings $\kappa_V$ (for the tree-level $hVV$ coupling), $\kappa_f$ (for the $hf\bar{f}$ coupling) have been estimated, as well as $\kappa_\gamma$, $\kappa_{Z\gamma}$ and the Higgs self-coupling. These studies incorporate projected measurements from CLIC \cite{Abramowicz:2016zbo,Robson:2018zje}, CEPC \cite{CEPCPhysicsStudyGroup:2022uwl}, ILC \cite{ILCInternationalDevelopmentTeam:2022izu}, FCC-ee \cite{Bernardi:2022hny}, and the muon collider ($\mu$C) \cite{deBlas:2022aow, Forslund:2022xjq}. 
The fits are primarily based on Higgs-pole observables, including Higgs-strahlung $\ell^+\ell^- \to Zh$ and vector-boson fusion (VBF) $\ell^+\ell^- \to \nu\nu h$, $\ell^+\ell^- h$ production followed by Higgs decays. In addition, projected electroweak precision observables from $Z$ pole studies are included, while beam polarization is exploited where available to improve the collider sensitivity.

At high-energy lepton colliders the VBF mechanism provides the largest single-Higgs production cross-section, yielding $O(10^6)$ Higgs bosons or more. Still,  rare Higgs decay channels probing loop-induced couplings suffer from branching ratios of $\mathcal{O}(10^{-3})$, rendering their precision measurement challenging~\cite{Forslund:2022xjq}. 
Similarly, at the HL-LHC and future $pp$ colliders, Higgs coupling measurements also rely directly on Higgs production on-shell and decay. With the large rates that $pp$ machines can achieve through gluon fusion, the rare decay channels can be studied to some precision at HL-LHC. The expected precision scales approximately as $1/\sqrt{\mathcal{L}}$, with the HL-LHC producing over $10^8$ Higgs bosons, thus reaching sub-10\% precision for the $Z\gamma$ channel and FCC-hh possibly reaching sub-1\% with its full luminosity~\cite{deBlas:2025gyz}.
In this context, the next major improvement beyond the HL-LHC is expected to come from FCC-hh, which will extend and significantly improve the corresponding measurements performed at the LHC and HL-LHC. This motivates the exploration of alternative processes that probe the same EFT operators.

%%%%%%%%%%%%%%%%%%%%%%%%%%%%%%%%%%%%%%%%%%%%%%%%%%%%%%%%%%%%%%%%%%%%%%%%%%%%%%%%%%%%%%%%%%%%%%%%%%%%
\section{Diboson processes}
\label{sec:diboson}
%%%%%%%%%%%%%%%%%%%%%%%%%%%%%%%%%%%%%%%%%%%%%%%%%%%%%%%%%%%%%%%%%%%%%%%%%%%%%%%%%%%%%%%%%%%%%%%%%%%%

Sticking to exactly the same effective interaction that modifies $h\to Z \gamma$ and $h\to \gamma\gamma$ we now study the annihilation of fermions into a $Vh$ or $VV$ final state
\begin{eqnarray}
    \ell^+ \ell^- \to V h, VV \label{eq:2body-signal}\,,
\end{eqnarray}
with $V = \gamma, Z, W$.
These processes have the advantage over Higgs decays of probing energy scales as high as the center-of-mass energy of the $\ell^+\ell^-$ annihilation process, which can greatly exceed the Higgs boson mass and thereby enhance the effects of the effective interaction. We recall that scattering amplitudes induced by
EFT couplings of dimension six, as those of \cref{eq:opHVVSMEFT}, need to carry an energy dependence $E^2/\Lambda^2$, where $E$ is a typical energy scale of the process, and $\Lambda$ is the mass scale of the EFT couplings.
This energy scale appearing in the cross-sections can be a fixed scale for processes at the Higgs pole, e.g.\ a factor $m_h^2/\Lambda^2$ for each EFT vertex in the decay amplitude, or a kinematic variable such as the transverse momentum of the final-state particles in the processes of \cref{eq:2body-signal}, thus resulting in $p_{T,h}^2/\Lambda^2$ or similar factors.
In high energy collisions at $\sqrt{s} \gg m_h$ we can reach kinematic regimes in which
\begin{eqnarray}
    \frac{m_h^2}{\Lambda^2} \ll  \frac{p_{T,h}^2}{\Lambda^2},
\end{eqnarray}
thus the effect of EFT interactions can be greatly magnified compared to pole observables. 
This effect has been exploited in~\cite{Buttazzo:2020uzc} to show that the EFT operators $\mathcal{O}_B$ and $\mathcal{O}_W$ can be constrained from diboson production at the 10~TeV muon collider, as well as in~\cite{Franceschini:2017xkh} for the HL-LHC. These constraints can be comparable, and under some assumptions even stronger, than the ones attainable at low-energy precision machines such as LEP or even FCC-ee.

A high-energy lepton collider with a center-of-mass energy of $E_{\rm cm}\approx 10\,\text{TeV}$ is expected to collect an integrated luminosity that scales with energy as
\begin{equation}\label{eq:lumi-scaling}
\mathcal{L}=10\,{\rm ab}^{-1}\left(\frac{E_{\rm cm}}{10\,{\rm TeV}}\right)^2\,.
\end{equation}
As such, the cross sections of the processes mentioned above could be measured with a precision at the few-percent level~\cite{Buttazzo:2020uzc,Costantini:2020stv,Holmes:2025ybv}.
At face value this might look like it entails a precision on involved couplings at a similar level of precision. However, the effects of higher-dimensional operators are expected to grow as the energy increases, so that the sensitivity to new physics of these processes could be enhanced with respect to the precision on the rate itself. 

Keeping this fact in mind, we start our analysis of high-energy scatterings with the diboson processes $\ell^+\ell^-\to VV$ and $\ell^+\ell^-\to Vh$, where $V = \gamma, Z, W^\pm$. % 
These processes are known to provide extremely high sensitivities to dimension-6 operators that involve Higgs currents~\cite{Buttazzo:2020uzc}. Here we extend the same philosophy to the analysis to the interactions \cref{eq:opHVVSMEFT}.

%%%%%%%%%%%%%%%%%%%%%%%%%%%%%%%%%%%%%%%%%%%%%%%%%%%%%%%%%%%%%%%%%%%%%%%%%%%%%%%%%%%%%%%%%%%%%%%%%%%%
\subsection{$\ell^+\ell^- \to VV$}
\label{sec:VV}

%%%%%%%%%%%%%%%%%%%%%%%%%%%%%%%%%%%%%%%%%%%%%%%%%%%%%%%%%%%%%%%%%%%%%%%%%%%%%%%%%%%%%%%%%%%%%%%%%%%%
\begin{figure}[t]
    \centering
        \includegraphics[width=0.36 \textwidth ]{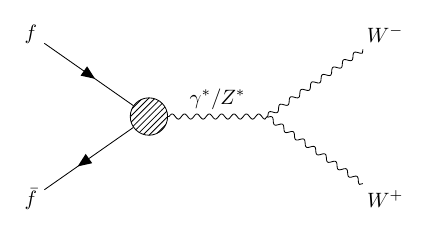}\hfill
        \includegraphics[width=0.36 \textwidth ]{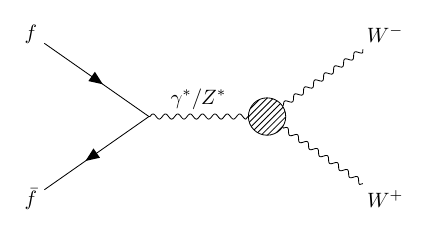}\hfill
        \includegraphics[width=0.27 \textwidth ]{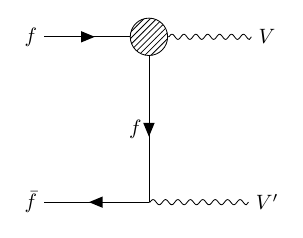}
        \caption{NP contributions to $VV$ production. For $W^+W^-$ only the s-channel diagrams contribute, for $ZZ,Z\gamma,\gamma\gamma$ only the $t$-channel contributes. Blobs represent the EFT interaction $\mathcal{O}_{HWB}$. \label{fig:VVdiagrams}}
\end{figure}
%%%%%%%%%%%%%%%%%%%%%%%%%%%%%%%%%%%%%%%%%%%%%%%%%%%%%%%%%%%%%%%%%%%%%%%%%%%%%%%%%%%%%%%%%%%%%%%%%%%%

The operators $\cO_{HW}$ and $\cO_{HB}$ modify the Higgs boson interactions, but they do not affect the gauge boson interaction vertices with fermions or among themselves, as discussed in \Cref{sec:ewpt}. Therefore, these two operators can contribute to $\ell^+\ell^-\to VV$ processes at tree-level only through the exchange of a Higgs boson in the $s$-channel, which is suppressed by the small lepton Yukawa coupling and will be neglected.

In contrast, the operator $\cO_{HWB}$ modifies both the triple gauge vertices $ZWW$, $\gamma WW$, and the fermion gauge couplings $Z\bar{f}f$, $\gamma\bar{f}f$. These modifications induce NP contributions to $\ell^+\ell^-\to VV$ proportional to $c_{HWB}$, through the diagrams shown in \Cref{fig:VVdiagrams}.

%%%%%%%%%%%%%%%%%%%%%%%%%%%%%%%%%%%%%%%%%%%%%%%%%%%%%%%%%%%%%%%%%%%%%%%%%%%%%%%%%%%%%%%%%%%%%%%%%%%%
\begin{figure}[t]
    \centering
    \includegraphics[width=0.49\linewidth]{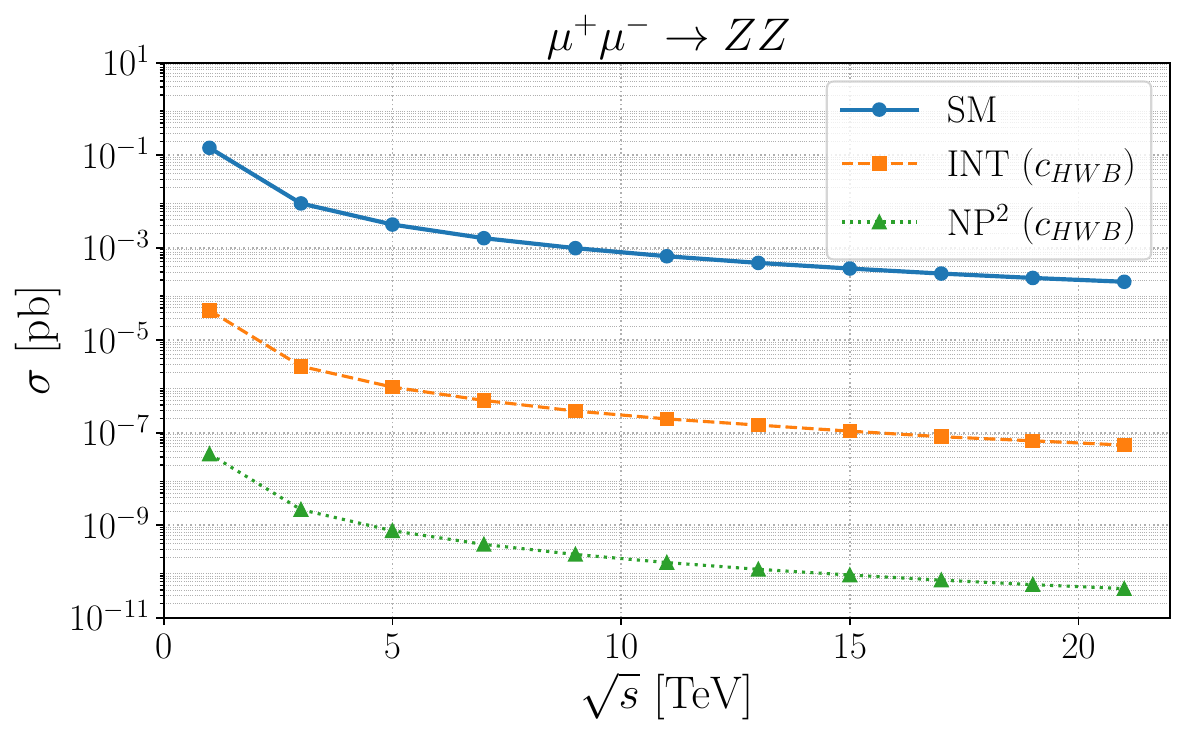}\hfill%
    \includegraphics[width=0.49\linewidth]{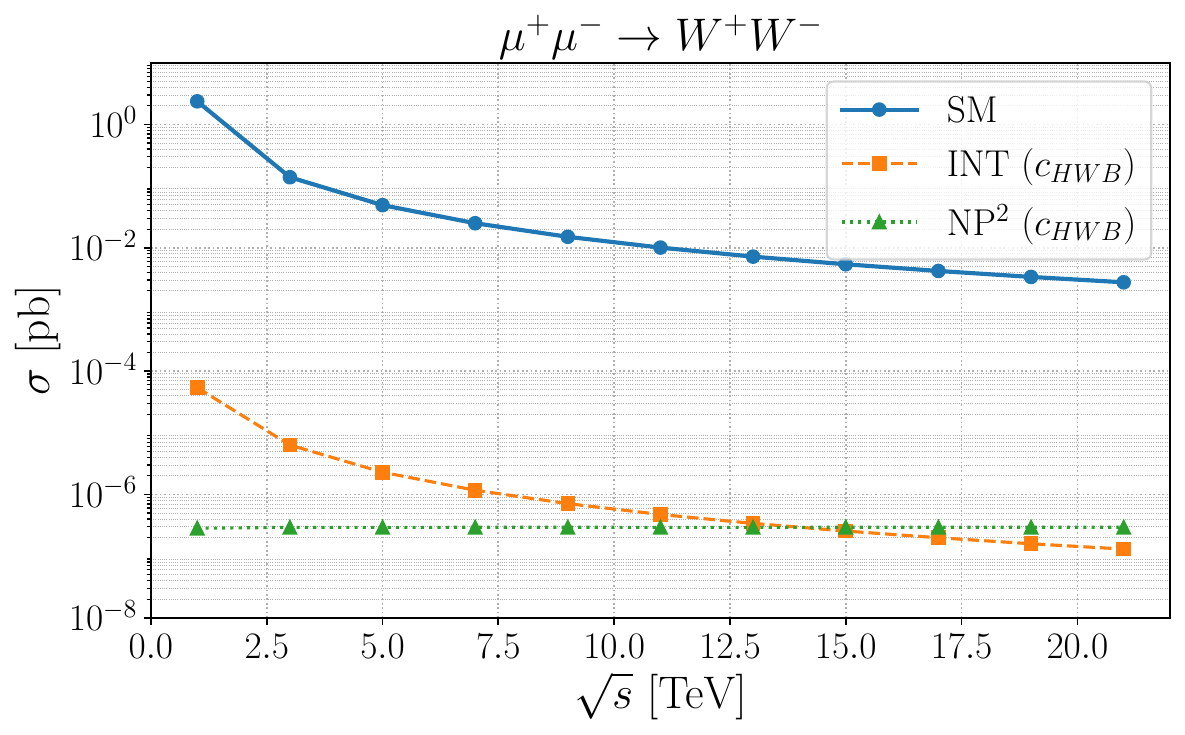}
    \caption{Cross sections for the processes $\ell^+ \ell^- \to Z Z$ (left) and $\ell^+ \ell^- \to W^+ W^-$ (right) plotted as a function of collider center-of-mass energy, for $\Lambda=20$ TeV and $c_{HWB}=1$. The contributions from SM and NP amplitudes, and their interference, are shown separately.}
    \label{fig:mupmum_zz}
\end{figure}
%%%%%%%%%%%%%%%%%%%%%%%%%%%%%%%%%%%%%%%%%%%%%%%%%%%%%%%%%%%%%%%%%%%%%%%%%%%%%%%%%%%%%%%%%%%%%%%%%%%%

The triple gauge couplings among neutral vectors $Z,\gamma$ vanish. Therefore, the neutral diboson processes $\ell^+\ell^-\to ZZ,\gamma\gamma,Z\gamma$ will only get corrections from the $t$-channel diagrams with a modified fermion gauge coupling in one of the vertices. Both the SM and NP cross-sections depend on renormalizable couplings and for a $2 \to 2$ scattering they scale with the center-of-mass energy as $E_{\rm cm}^{-2}$.
For the neutral channels the NP amplitudes are thus proportional to the SM ones,
\begin{equation}\label{eq:AZZ}
    \Amp_{ZZ} \approx c_{HWB}\frac{v^2}{\Lambda^2}\Amp_{ZZ}^{\rm SM},\qquad\qquad \text{(and similarly for $\gamma\gamma$, $Z\gamma$)}
\end{equation}
and indeed have the same energy dependence.
This can be clearly seen in \Cref{fig:mupmum_zz} (left panel) for the $ZZ$ final state. The SM amplitude, its interference with NP, and the NP squared amplitudes all give rise to cross-sections that fall with the same power of the center-of-mass energy $E_{\rm cm}^{-2}$.

This should be contrasted with what happens in the $W^+W^-$ channel. The charged gauge coupling with fermions is not modified by $c_{HWB}$ in the $m_W$ scheme~\cite{Brivio:2017bnu}, therefore the charged process $\ell^+\ell^-\to W^+W^-$ gets a NP contribution only from the exchange of $Z,\gamma$ in the $s$-channel.
The $ZWW$ and $\gamma WW$ triple gauge vertices are modified due to the EFT interactions as follows,
\begin{equation}
    \L_{WWV} \supset -i g_{WWV} (1+\delta g_1^V)\big(W_{\mu\nu}^+W^{-\mu}V^\nu - W_{\mu\nu}^-W^{+\mu}V^\nu\big)\,,
\end{equation}
where $V = Z,\gamma$, $g_{WW\gamma} = e$, and $g_{WWZ} = e \cot\theta_w$, and~\cite{Alonso:2013hga}
\begin{align}
\delta g_1^\gamma 
&= -\frac{1}{\sqrt{2}{G}_F}
 \frac{ {m}_W}{\sqrt{{m}_Z^2 - m_W^2}}\frac{c_{HWB}}{\Lambda^2}\,,
&
\delta g_1^Z 
&= \frac{1}{\sqrt{2} G_F}
\frac{m_Z}{m_W}
\sqrt{1 - \frac{{m}_W^2}{m_Z^2}} \, \frac{c_{HWB}}{\Lambda^2}\,.
\end{align}
The dominant contribution arises when one of the final-state vectors has longitudinal polarization, while the other one is transverse, leading to an enhancement of the NP amplitude at high energies,
\begin{equation}\label{eq:AWW}
    \Amp_{WW} \approx c_{HWB}\frac{v E_{\rm cm}}{\Lambda^2}\Amp_{WW}^{\rm SM},
\end{equation}
enhanced by a factor $E_{\rm cm}/v$ with respect to \cref{eq:AZZ}. The energy dependence of the $\ell^+\ell^-\to W^+W^-$ cross-section is shown in the right panel of \Cref{fig:mupmum_zz}. The NP contribution remains constant with increasing $E_{\rm cm}$, unlike in the $ZZ$ case, while the interference scales as $1/E_{\rm cm}^2$ like the SM contribution.

The dependence on energy of the $VV$ processes displays the two possible routes to attain information on the coefficient $c_{HWB}$. For the neutral final states with no energy enhancements there is no gain in signal-to-noise ratio when increasing the center-of-mass energy, and one expects a similar precision as from total rate measurements at low energies. On the contrary, for $WW$ the signal-to-noise ratio increases at higher energies, hence the effort to increase   the center-of-mass energy is rewarded with a better sensitivity to the Wilson coefficient.

%%%%%%%%%%%%%%%%%%%%%%%%%%%%%%%%%%%%%%%%%%%%%%%%%%%%%%%%%%%%%%%%%%%%%%%%%%%%%%%%%%%%%%%%%%%%%%%%%%%%
\begin{figure}[t]
    \centering
    \includegraphics[width=0.65\linewidth]{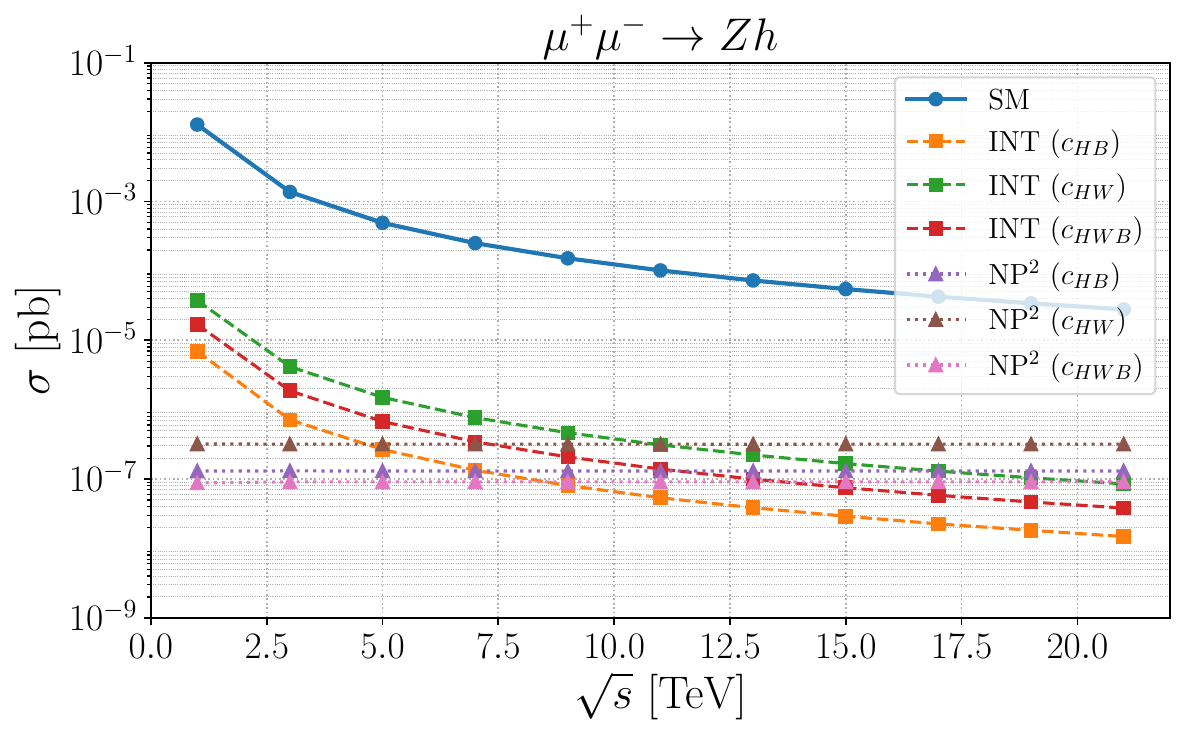}
    \caption{Contributions to the $\ell^+ \ell^- \to Z h$ cross-section as a function of center-of-mass energy, for $\Lambda=20$ TeV and $c_{i}=1$.}
    \label{fig:mupmumZh}
\end{figure}
%%%%%%%%%%%%%%%%%%%%%%%%%%%%%%%%%%%%%%%%%%%%%%%%%%%%%%%%%%%%%%%%%%%%%%%%%%%%%%%%%%%%%%%%%%%%%%%%%%%%

%%%%%%%%%%%%%%%%%%%%%%%%%%%%%%%%%%%%%%%%%%%%%%%%%%%%%%%%%%%%%%%%%%%%%%%%%%%%%%%%%%%%%%%%%%%%%%%%%%%%
\subsection{$\ell^+\ell^- \to Vh $}
\label{sec:Vh}

The processes $\ell^+\ell^-\to Zh$ and $\ell^+\ell^-\to \gamma h$ are directly affected by all three operators $\cO_{HW}$, $\cO_{HB}$, and $\cO_{HWB}$, through the modified $VVh$ couplings $\kappa_\gamma$, $\kappa_{\gamma Z}$ that they induce.

The SM cross-section, as usual at lepton colliders, falls with the center-of-mass energy $E_{\rm cm}$, owing to the $m_Z/E_{\rm cm}$ suppressed production of transverse $Z$ bosons. The pure NP cross-sections, on the contrary, are constant with respect to center-of-mass energy, as expected from the momentum dependence originating from the NP vertices.  \Cref{fig:mupmumZh} displays all the $Vh$ rates of our interest separating SM, NP and interference contributions. 
Note that, neglecting the muon Yukawa coupling, the final state $h \gamma$ is produced exclusively at one loop in the SM, and is thus suppressed, favoring the dominance of NP contributions.

%%%%%%%%%%%%%%%%%%%%%%%%%%%%%%%%%%%%%%%%%%%%%%%%%%%%%%%%%%%%%%%%%%%%%%%%%%%%%%%%%%%%%%%%%%%%%%%%%%%%
\begin{figure}[t]
    \centering%
    \includegraphics[width=0.32\textwidth]{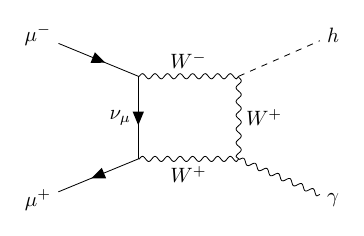}
    \hfill
    \includegraphics[width=0.33\textwidth]{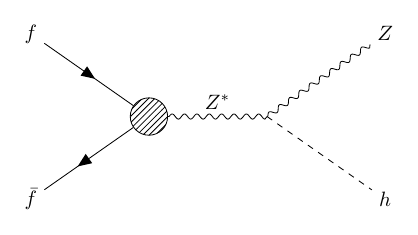}
    \hfill
    \includegraphics[width=0.33\textwidth]{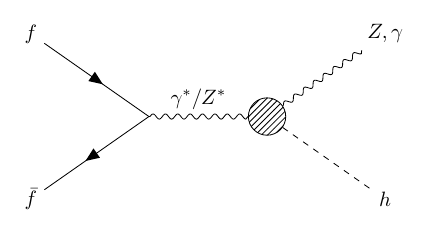}
    \caption{SM and EFT contributions to $h\gamma$ and $hZ$ production.
    Blobs represent EFT interactions of \cref{eq:opHVVSMEFT}.
    The box diagram is representative of the loop-induced SM contribution
    and does not show all SM diagrams contributing to the process.}
    \label{fig:Vhdiagrams}
\end{figure}
%%%%%%%%%%%%%%%%%%%%%%%%%%%%%%%%%%%%%%%%%%%%%%%%%%%%%%%%%%%%%%%%%%%%%%%%%%%%%%%%%%%%%%%%%%%%%%%%%%%%

%%%%%%%%%%%%%%%%%%%%%%%%%%%%%%%%%%%%%%%%%%%%%%%%%%%%%%%%%%%%%%%%%%%%%%%%%%%%%%%%%%%%%%%%%%%%%%%%%%%%
\subsection{Constraints from dibosons}\label{sec:diboson-summary}

To compute cross-section and derive bounds  we use the SMEFTsim~\cite{Brivio:2020onw} \texttt{FeynRules} \cite{Alloul:2013bka} model with the appropriate $\{m_W, m_Z, G_F\}$ input scheme defined for the SM parameters. Therefore, the model takes into account the aforementioned vertex modifications due to NP presence automatically. The \texttt{UFO} model is then imported to \texttt{MG5\_aMC@NLO} \cite{Alwall:2014hca} where we calculate the cross-sections with varying beam energy and no PDF convolution, as appropriate for elementary beam particles such as muons. We identify for every process also the dominant background contributions and confront it with the signal as a function of the relevant Wilson coefficients.

All bosons are assumed to decay hadronically. As a source of background we also consider the fact that, due to final di-jet mass resolution $\Delta M_{jj}$, it is possible to misidentify a Higgs boson as a $Z$ or vice-versa. This for example requires us to include $\ell^+ \ell^- \to  Z\gamma$ as a source of background for the process $\ell^+ \ell^- \to h \gamma$. For the remainder of this work, we set $\Delta M_{jj}/M_{jj} = 5\% $, which we take as a realistic scenario accessible at very high energy lepton colliders in the future~\cite{Holmes:2025ybv,MAIA:2025hzm}. 
To obtain an estimate for the efficiencies, we model the invariant mass distributions of the signal and background processes as Gaussians smeared by the mass resolution $\Delta M_{jj}/M_{jj}$. Applying a mass-window cut on the final states proves to be very effective in reducing the misidentified background, while retaining almost all of the signal events. %  
We also apply an angular cut on one of the outgoing particles where it may help in separating the NP signal from the SM background.
In particular, for the final state $W^+W^-$ the angular cut significantly suppresses the SM background, which receives its dominant contribution from forward-enhanced $t$-channel production of transversely polarized $W$ bosons. The invariant mass cuts and the angular cuts are mentioned for all the processes in \Cref{tab:cuts}. In addition, we impose acceptance cuts $p_T > 10$~GeV, $\eta < 2.5$ for all particles.
The obtained cross-sections are further multiplied with the branching ratios, according to the experimental signature of the signal.
As an example, for the final state $h\gamma$ (with background $Z\gamma$) we consider the decay of $h \to b\bar{b}$ ($Z \to b\bar{b}$) along with the probability of tagging at least 1 $b$-jet. The $b$-tagging efficiency that we used was $\varepsilon_b = 70\%$. 

%%%%%%%%%%%%%%%%%%%%%%%%%%%%%%%%%%%%%%%%%%%%%%%%%%%%%%%%%%%%%%%%%%%%%%%%%%%%%%%%%%%%%%%%%%%%%%%%%%%%
\begin{figure}[t]
    \centering%
\includegraphics[width = 0.5\textwidth]{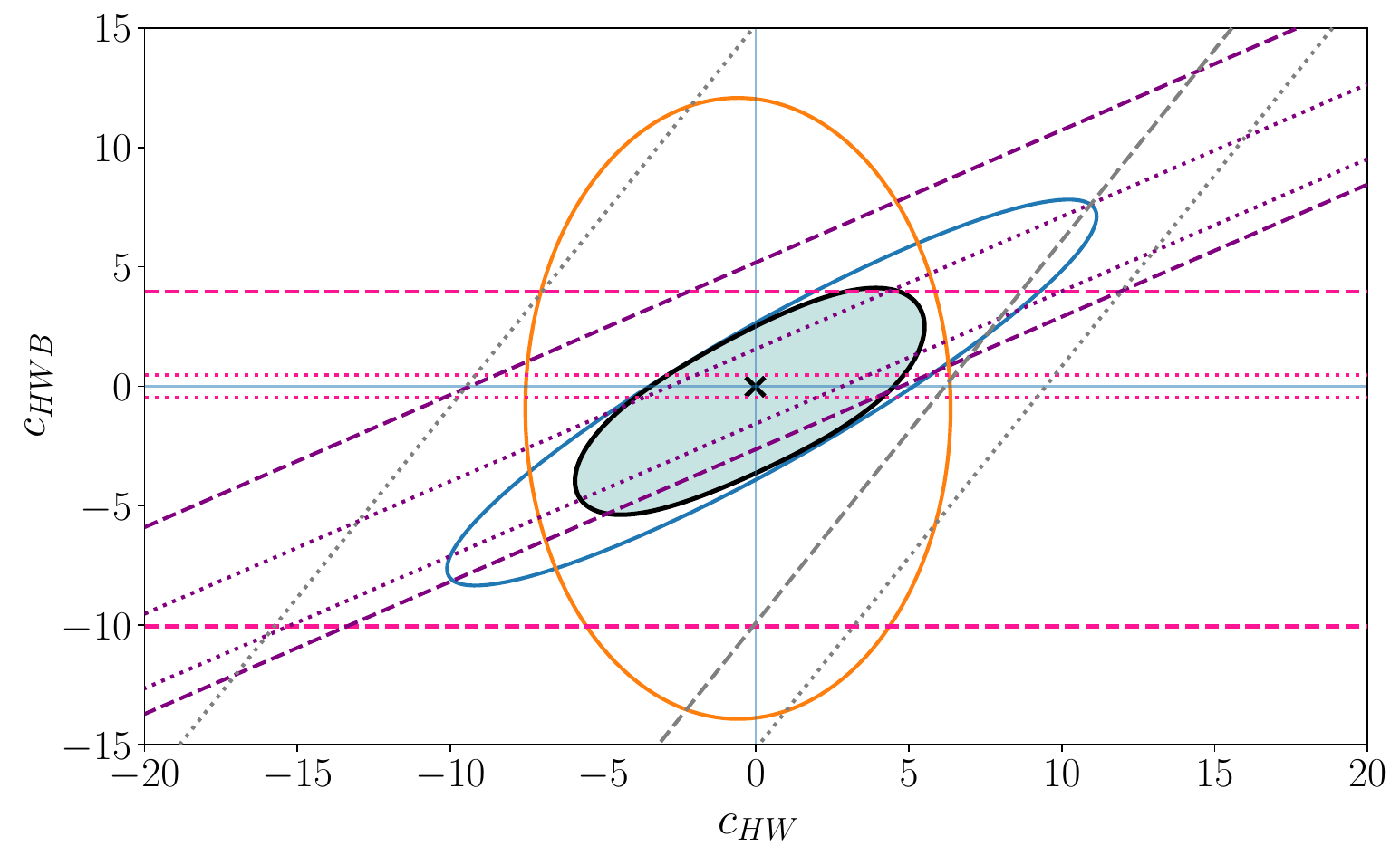}\hfill%
\includegraphics[width = 0.5\textwidth]{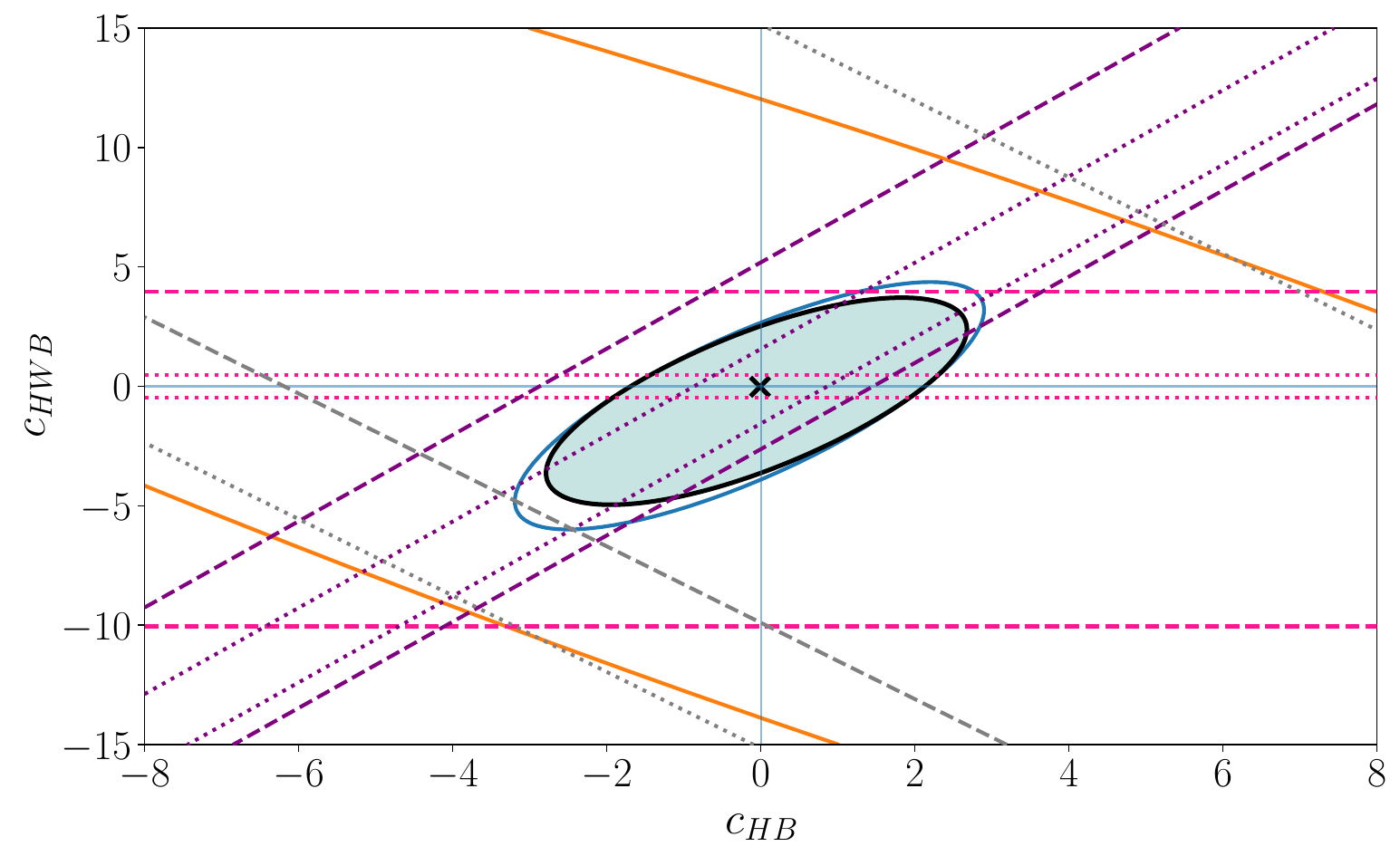}\\
\raisebox{0.1\height}{\includegraphics[width=0.43\textwidth]{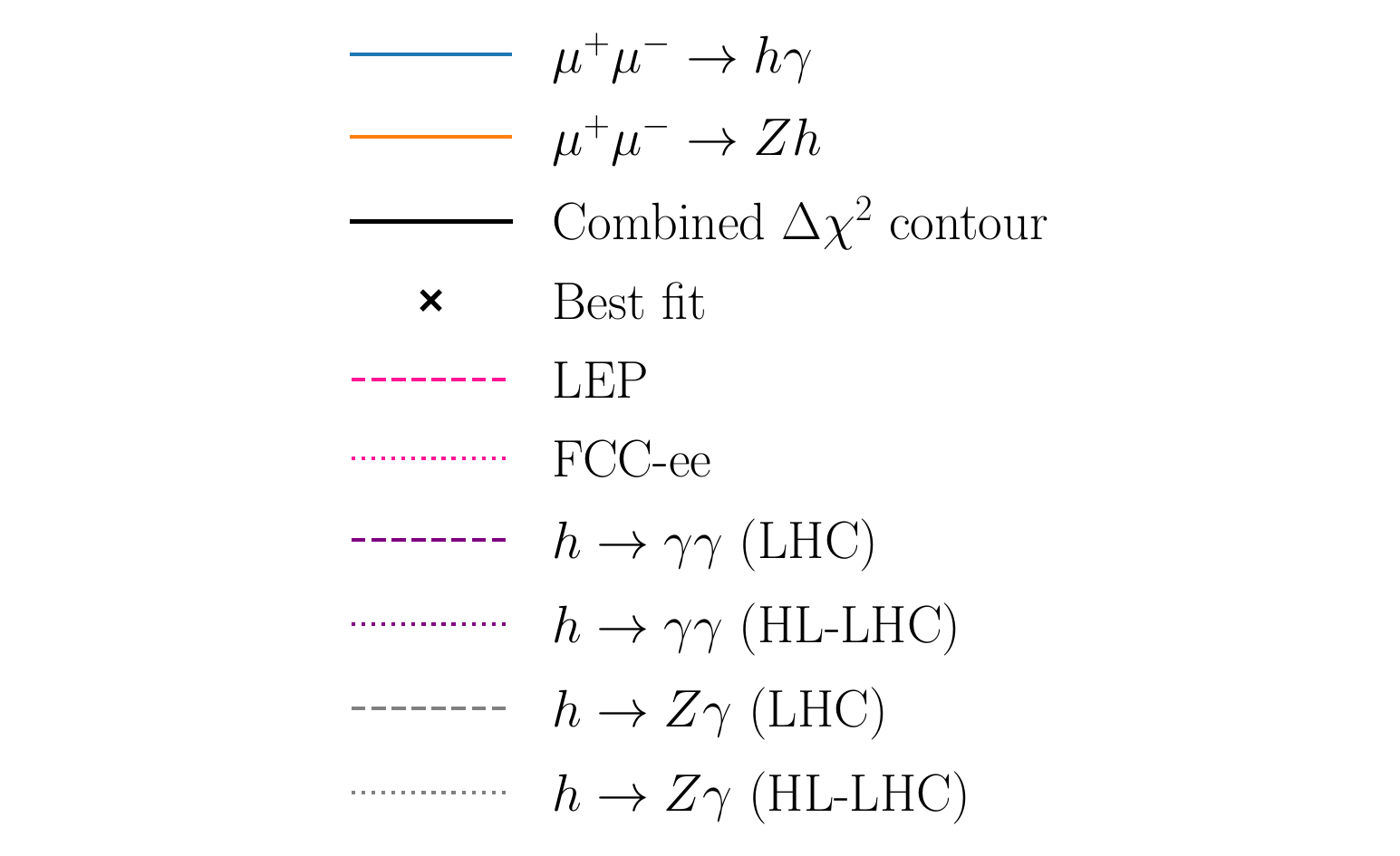}}\hfill%
\includegraphics[width=0.5\textwidth]{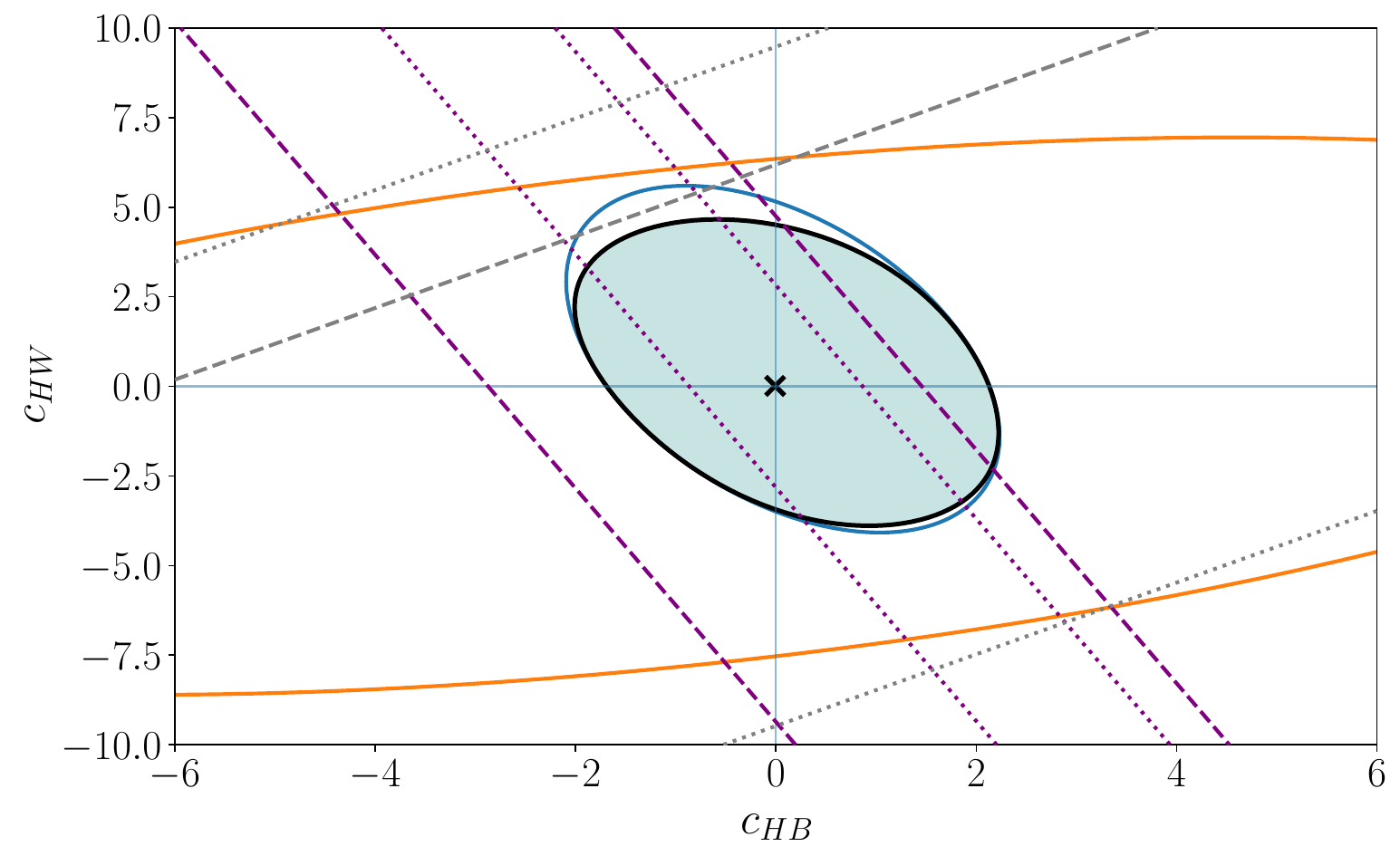}
    \caption{Results of the $\chi^2$ fit for diboson processes at $\sqrt{s} = 10$~TeV in the $(c_{HB}, c_{HW})$, $(c_{HB}, c_{HWB})$, and $(c_{HW}, c_{HWB})$ planes. The third Wilson coefficient in each panel is set to zero, and the NP scale is normalized to $\Lambda = 20$~TeV. The filled regions represent the combined fit from all $\mu^+ \mu^- \to VV$ and $\mu^+ \mu^- \to Vh$ diboson processes.}
    \label{fig:2to2}
\end{figure}
%%%%%%%%%%%%%%%%%%%%%%%%%%%%%%%%%%%%%%%%%%%%%%%%%%%%%%%%%%%%%%%%%%%%%%%%%%%%%%%%%%%%%%%%%%%%%%%%%%%%

\Cref{fig:2to2} shows the two-dimensional 95\% CL bounds on the SMEFT Wilson coefficients. The contours are obtained from $\chi^2$ fits in the plane of two coefficients, while the third coefficient is fixed to zero.
The uncertainties were assumed to be Poissonian, associated to the expected number of total (signal plus background) events. The center-of-mass energy is taken to be $\sqrt{s} = 10\,\mathrm{TeV} $, and the NP scale was set to $\Lambda = 20\,\mathrm{TeV}$. We also overlay the bounds on the Wilson coefficient $c_{HWB}$ derived from current and projected  EWPO constraints. These are taken from  \cref{eq:EWPOcurrent} and \cref{eq:EWPOprojected}, respectively. The current (LHC) and future (HL-LHC) bounds derived from the rare Higgs decays $h \to \gamma \gamma$, $h \to Z \gamma$  are also shown in the panels.

From the results, it is evident that $\ell^+ \ell^- \to h\gamma$ provides the strongest sensitivity, mainly due to the loop suppression of the corresponding SM irreducibile background \cite{Abbasabadi:1995rc,Djouadi:1996ws}, combined with the high efficiency of the applied invariant mass cut. The SM $h\gamma$ background has been included in our analysis; however, we have verified that its contribution has a negligible impact on the final results. We remark that, although the projected bounds from $h\to\gamma\gamma$ remain stronger than what we can obtain from diboson processes, our analysis significantly improves the constraints in the direction primarily probed by $h\to Z\gamma$.

%%%%%%%%%%%%%%%%%%%%%%%%%%%%%%%%%%%%%%%%%%%%%%%%%%%%%%%%%%%%%%%%%%%%%%%%%%%%%%%%%%%%%%%%%%%%%%%%%%%%
\subsection{EFT validity}\label{sec:EFTvalidity}

%%%%%%%%%%%%%%%%%%%%%%%%%%%%%%%%%%%%%%%%%%%%%%%%%%%%%%%%%%%%%%%%%%%%%%%%%%%%%%%%%%%%%%%%%%%%%%%%%%%%
\begin{figure}[t!]
    \centering
    \includegraphics[width=0.75\linewidth]{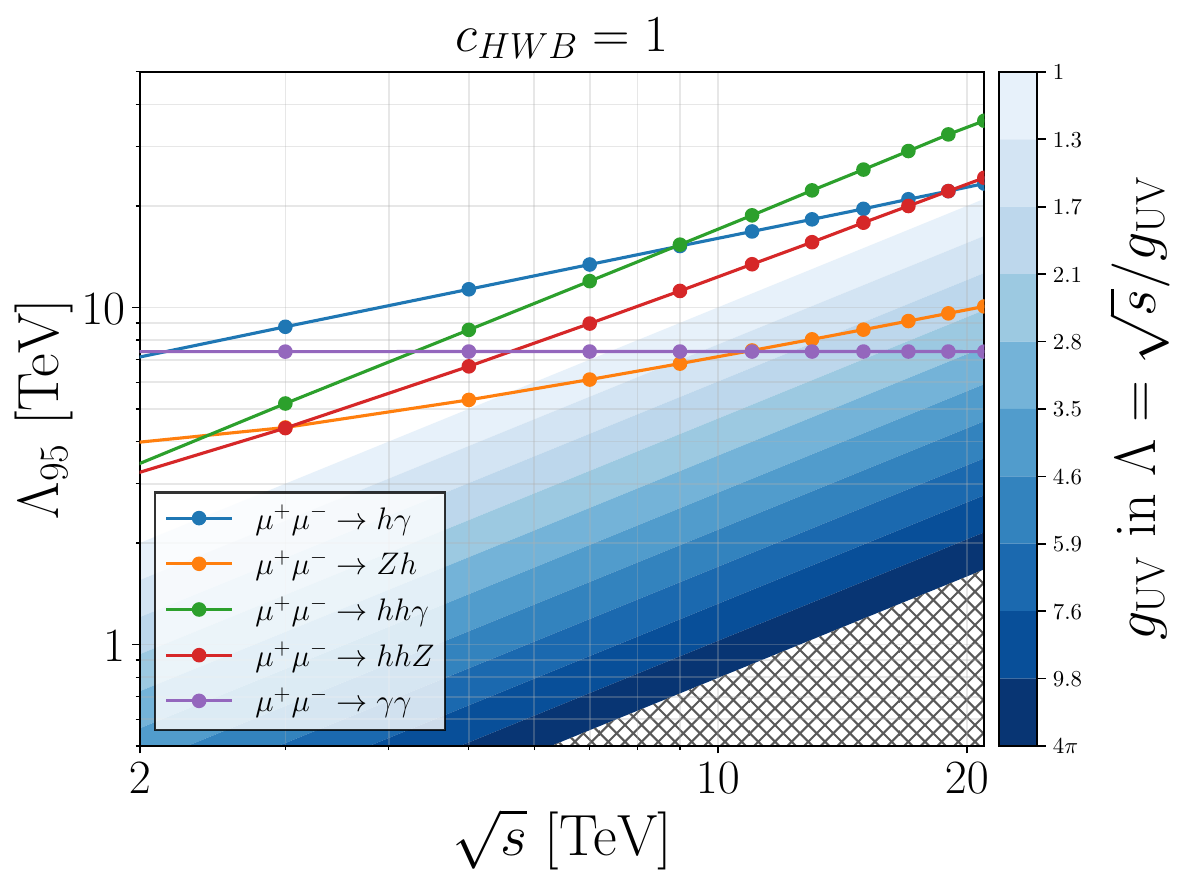}
    \caption{Lower bound on the EFT scale $\Lambda_{95}$ at 95\% CL, as a function of the c.o.m. energy $\sqrt{s}$ for a set of processes. Only the operator $\mathcal{O}_{HWB}$ is included, with Wilson coefficient $c_{HWB} = 1$, and the remaining two Wilson coefficients are set to zero. The blue shadings show contours of constant NP coupling $g_{\rm UV}\equiv \sqrt{s}/\Lambda$.}
    \label{fig:twototwo-and-three-eft-validity}
\end{figure}
%%%%%%%%%%%%%%%%%%%%%%%%%%%%%%%%%%%%%%%%%%%%%%%%%%%%%%%%%%%%%%%%%%%%%%%%%%%%%%%%%%%%%%%%%%%%%%%%%%%%

From the discussion above, it is evident that our analysis of two-boson final states may incur in the problem of involving transferred momentum larger than the cut-off of the EFT, defined as the scale where new particles of mass $M_{\rm NP}$ can be produced on-shell. Such particles exist in the UV models of NP, but are absent in the EFT at lower energy, in which our computations were carried out.
Quantitatively, the EFT cut-off is related to the scale\footnote{Here and in the following of this discussion we set $c_{i} = 1$ for simplicity. The physical scale should always be understood as $\Lambda/\sqrt{c_{i}}$.} $\Lambda$ appearing in \cref{eq:lagrHVVSMEFT} as 
$M_{\rm NP} = \Lambda \cdot g_{\rm NP} < 4\pi\Lambda$, where $g_{\rm NP}<4\pi$ is a representative coupling strength of the new physics. 
If the bound on $\Lambda$ is not sufficiently strong, we may violate the EFT assumption that the new particles with mass $M_{\rm NP}$ are too heavy to be produced on-shell, or nearly on-shell, in diboson production. As a simple condition for the validity of the EFT description, we impose $M_{\rm NP} > \sqrt{s}$.

In particular, the absence of NP effects that grow with energy lets us anticipate that increasing the c.o.m.\ energy will eventually cause the EFT expansion to break down. 
Indeed, the limit on the NP cross-section (in Gaussian approximation and neglecting systematic uncertainties) scales with the inverse square root of the integrated luminosity $\mathcal{L}$, 
\begin{equation}
    \sigma_{\rm NP}
   \lesssim \sqrt{\sigma_{\rm SM}/\mathcal{L}}\propto E_{\rm cm}^{-2},
\end{equation} 
where we used the energy dependence given in~\cref{eq:lumi-scaling} and $\sigma_{\rm SM}\propto E_{\rm cm}^{-2}$. Since $\sigma_{\rm NP}\propto \Lambda^{-4}$, one sees that if the NP cross-section grows slower than $E_{\rm cm}^2$, the limit on $\Lambda$ increases less than linearly with $E_{\rm cm}$, and will eventually violate the bound $\Lambda > E_{\rm cm}/4\pi$ at some energy.\footnote{Analogously, the interference between SM and NP is proportional to $\Lambda^{-2}$, thus it violates the EFT bound if it grows slower than a constant.}

\Cref{fig:twototwo-and-three-eft-validity} shows the values $\Lambda_{95}$ of the NP scale excluded at 95\% CL for different processes, considering only the operator $\mathcal{O}_{HWB}$. In particular, the bounds from $\ell^+\ell^-\to h\gamma$ (blue line) and $\ell^+\ell^-\to Zh$ (orange line) scale as $\Lambda_{95}\propto E_{\rm cm}^{1/2}$, consistently with the NP cross-section being constant for these processes. The constraint from $\ell^+\ell^-\to \gamma\gamma$ (purple line), for which the NP cross-section scales as $E_{\rm cm}^{-2}$, yields a limit on $\Lambda_{95}$ which is independent of the collider energy.
In the same plot, we also show contours of constant $g_{\rm UV} \equiv \sqrt{s}/\Lambda$, which is the minimal value of the NP coupling such that $M_{\rm NP} > \sqrt{s}$.

For center-of-mass energies $\sqrt{s}\lesssim 20$~TeV the EFT remains largely valid
for couplings $g_{\rm UV}\lesssim  1\ll 4\pi$. We remark that this analysis assumes the NP operators to be generated at tree-level. If they instead arise at one-loop in a concrete UV completion as the ones sketched out in \Cref{sec:EFT}, the values of $g_{\rm UV}$ in \Cref{fig:twototwo-and-three-eft-validity} need to be rescaled by a factor $4\pi$ that compensates the additional loop factor in the Wilson coefficients, see \cref{eq:power-counting}. In this case, only the results for $\ell^+\ell^-\to h\gamma$ (blue line in \Cref{fig:twototwo-and-three-eft-validity}) remain valid up to $\sqrt{s}\approx 20$~TeV for couplings $g_{\rm UV}^{\rm loop} < 4\pi$, while the other two considered diboson processes can test at most scales of a few TeV within the EFT.

%%%%%%%%%%%%%%%%%%%%%%%%%%%%%%%%%%%%%%%%%%%%%%%%%%%%%%%%%%%%%%%%%%%%%%%%%%%%%%%%%%%%%%%%%%%%%%%%%%%%
\section{Triboson processes}
\label{sec:triboson}
%%%%%%%%%%%%%%%%%%%%%%%%%%%%%%%%%%%%%%%%%%%%%%%%%%%%%%%%%%%%%%%%%%%%%%%%%%%%%%%%%%%%%%%%%%%%%%%%%%%%

%%%%%%%%%%%%%%%%%%%%%%%%%%%%%%%%%%%%%%%%%%%%%%%%%%%%%%%%%%%%%%%%%%%%%%%%%%%%%%%%%%%%%%%%%%%%%%%%%%%%
\begin{figure}[t]
    \centering
    \includegraphics[width=0.49\textwidth]{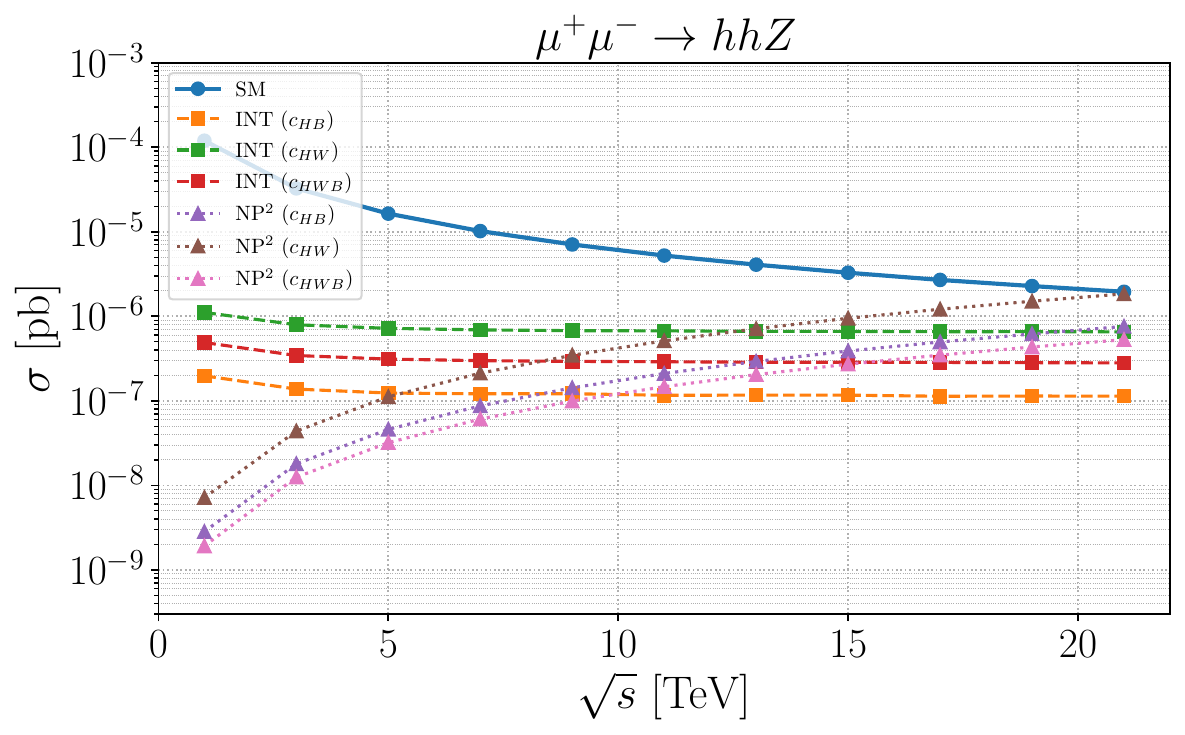}\hfill
    \includegraphics[width=0.49\textwidth]{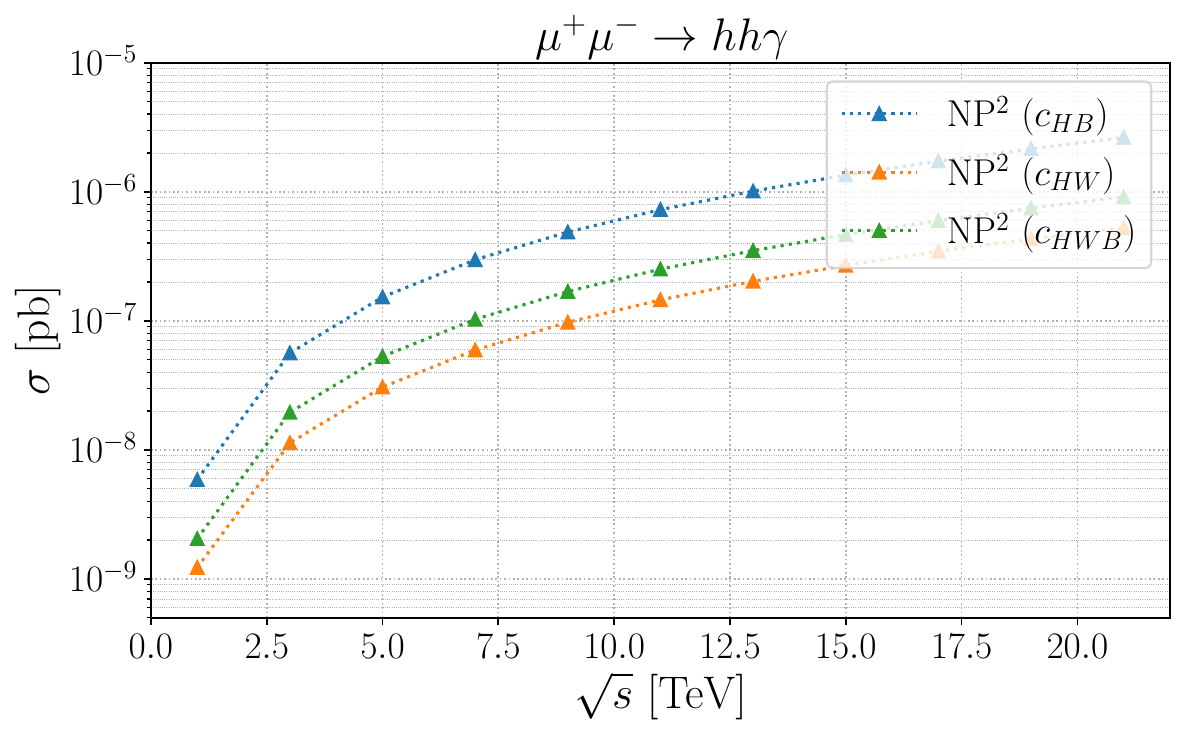}
    \caption{Cross section for the processes $\mu^+ \mu^- \to h h Z$ and $\mu^+ \mu^- \to h h \gamma$ plotted as a function of energy, for $\Lambda = 20$~TeV and $c_i = 1$.}
    \label{fig:mupmumHHZ}
\end{figure}
%%%%%%%%%%%%%%%%%%%%%%%%%%%%%%%%%%%%%%%%%%%%%%%%%%%%%%%%%%%%%%%%%%%%%%%%%%%%%%%%%%%%%%%%%%%%%%%%%%%%

The operators $\cO_{HWB},\cO_{HW},\cO_{HB}$ also contribute to multi-boson processes with a larger number of particles in the final state.
The rates for these processes are naively expected to be smaller than those for the corresponding two-body final states \cref{eq:2body-signal}, due to the additional phase-space suppression.
However, the scattering amplitudes for $2\to 2$ processes in \cref{eq:AWW} contain at least one Higgs vacuum expectation value (vev), which induces a suppression by $v/\Lambda$. On the contrary, the same operators contribute to $2\to 3$ processes without vev insertions, such that the NP amplitudes can grow as $\mathcal{A}_{\rm 3-body} \approx E_{\rm cm}^2/\Lambda^2$, with an enhancement by $E_{\rm cm}/v$ compared to the 2-body amplitudes of \cref{eq:AWW}. We can therefore anticipate that the three-body processes become the dominant constraint at sufficiently high scattering energies $E_{\rm cm} \gtrsim 4\pi v$.

For our study we consider the triboson processes
\begin{eqnarray}
      \ell^+ \ell^- &\to& WW\gamma, WWZ, ZZ\gamma, Z\gamma\gamma  ,   ZZZ  ,\\
       \ell^+ \ell^- &\to& hWW, hZ\gamma, h\gamma\gamma  ,\\
       \ell^+ \ell^- &\to& hh\gamma, hhZ  .
\end{eqnarray}
If these rates can be measured well enough at a multi-TeV muon collider, they can provide valuable information on the new physics under consideration, benefitting from the magnification of the sensitivity due to energy-growing effects.

We compute the signal and background rates using {\tt MG5\_aMC@NLO}~\cite{Alwall:2014hca} at parton level, as already detailed in \Cref{sec:diboson-summary}. The behaviours of the various contributions to the cross-section are given in the left panel of \Cref{fig:mupmumHHZ} as a function of center-of-mass energy for the example process $\ell^+\ell^-\to hhZ$. 
As expected, we can see that the NP cross-sections and the interference terms have a very different energy dependence with respect to the diboson case: there is an almost energy-independent contribution from the interference, and a contribution growing as $E_{\rm cm}^2$ from pure NP. 
This means that even when the signal mainly comes from the interference term there is an advantage in signal-to-noise ratio from going to higher center-of-mass energies. This differs from the diboson case, where only the pure NP terms benefit from an energy growth. The $\ell^+\ell^-\to hh\gamma$ cross-section, shown in the right panel of \Cref{fig:mupmumHHZ}, has a similar dependence on the Wilson coefficients and collider energy, with the caveat that the SM contribution is loop-suppressed in the limit of vanishing lepton Yukawa interaction, as in the case of $\ell^+\ell^-\to h\gamma$, hence only the pure NP contribution will be relevant.

%%%%%%%%%%%%%%%%%%%%%%%%%%%%%%%%%%%%%%%%%%%%%%%%%%%%%%%%%%%%%%%%%%%%%%%%%%%%%%%%%%%%%%%%%%%%%%%%%%%%
\begin{figure}[t]
    \centering%
\includegraphics[width = 0.5\linewidth]{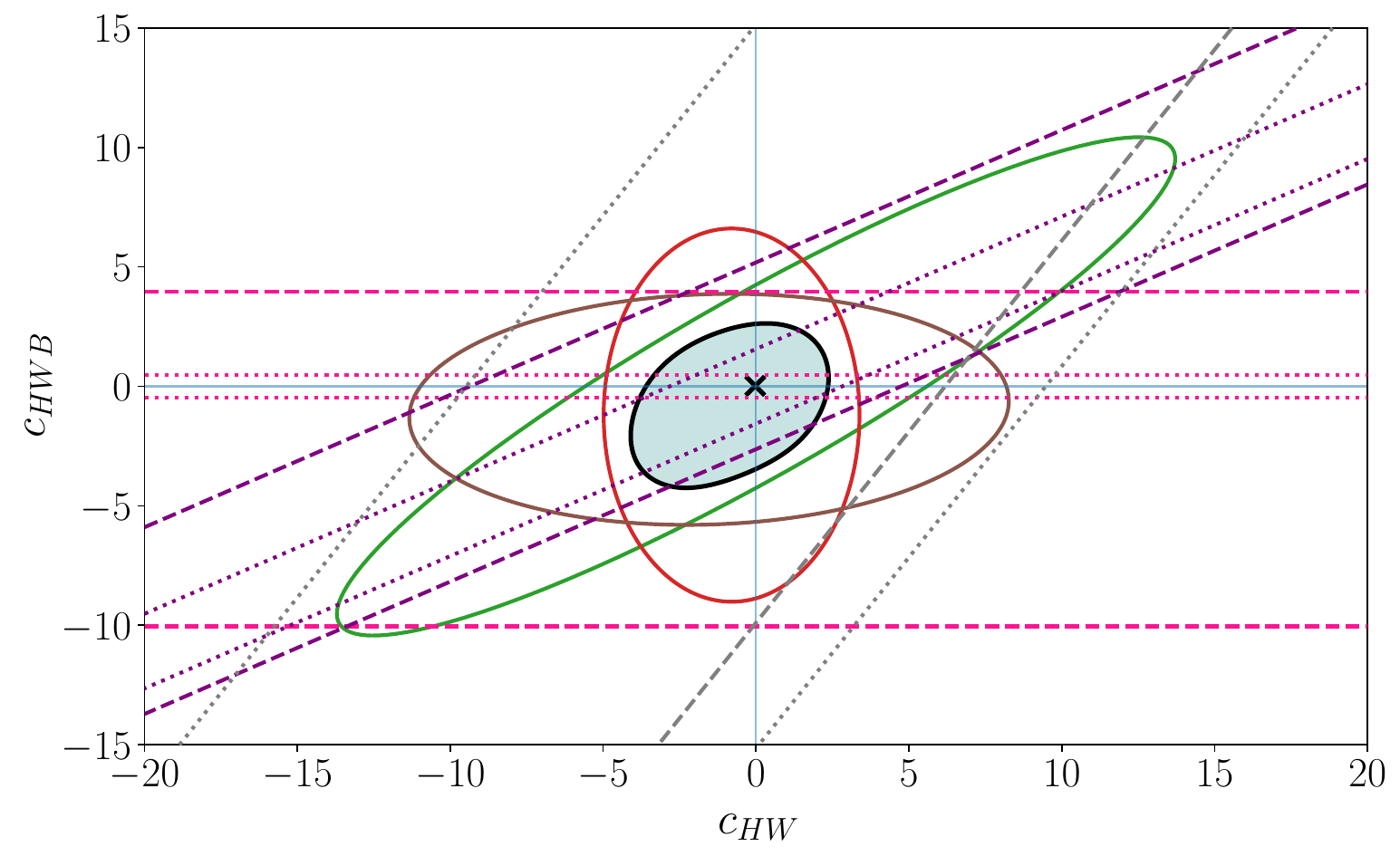}\hfill%
\includegraphics[width = 0.5\linewidth]{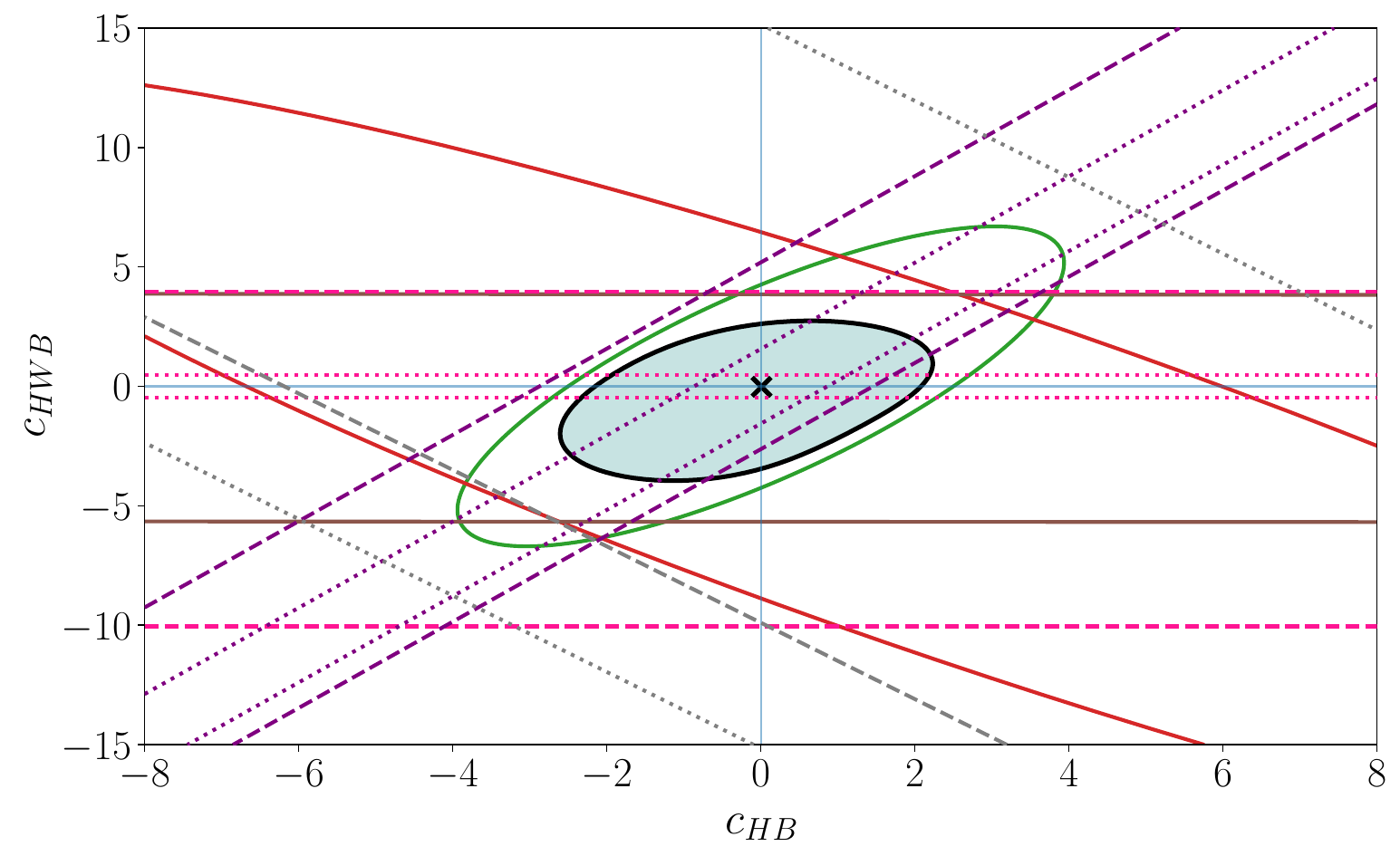}\\
\raisebox{0.05\height}{\includegraphics[width=0.48\textwidth]{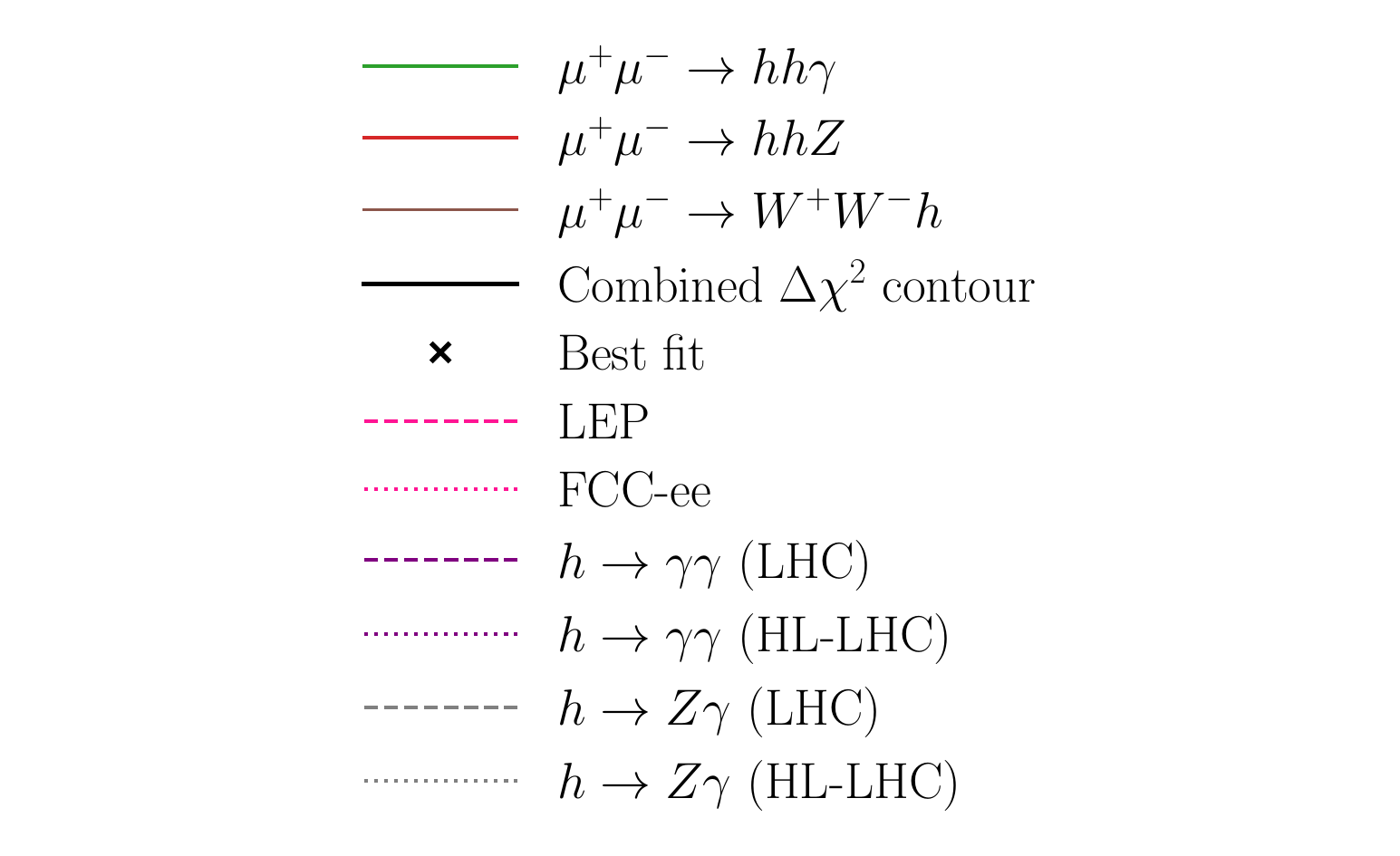}}\hfill%
\includegraphics[width=0.5\linewidth]{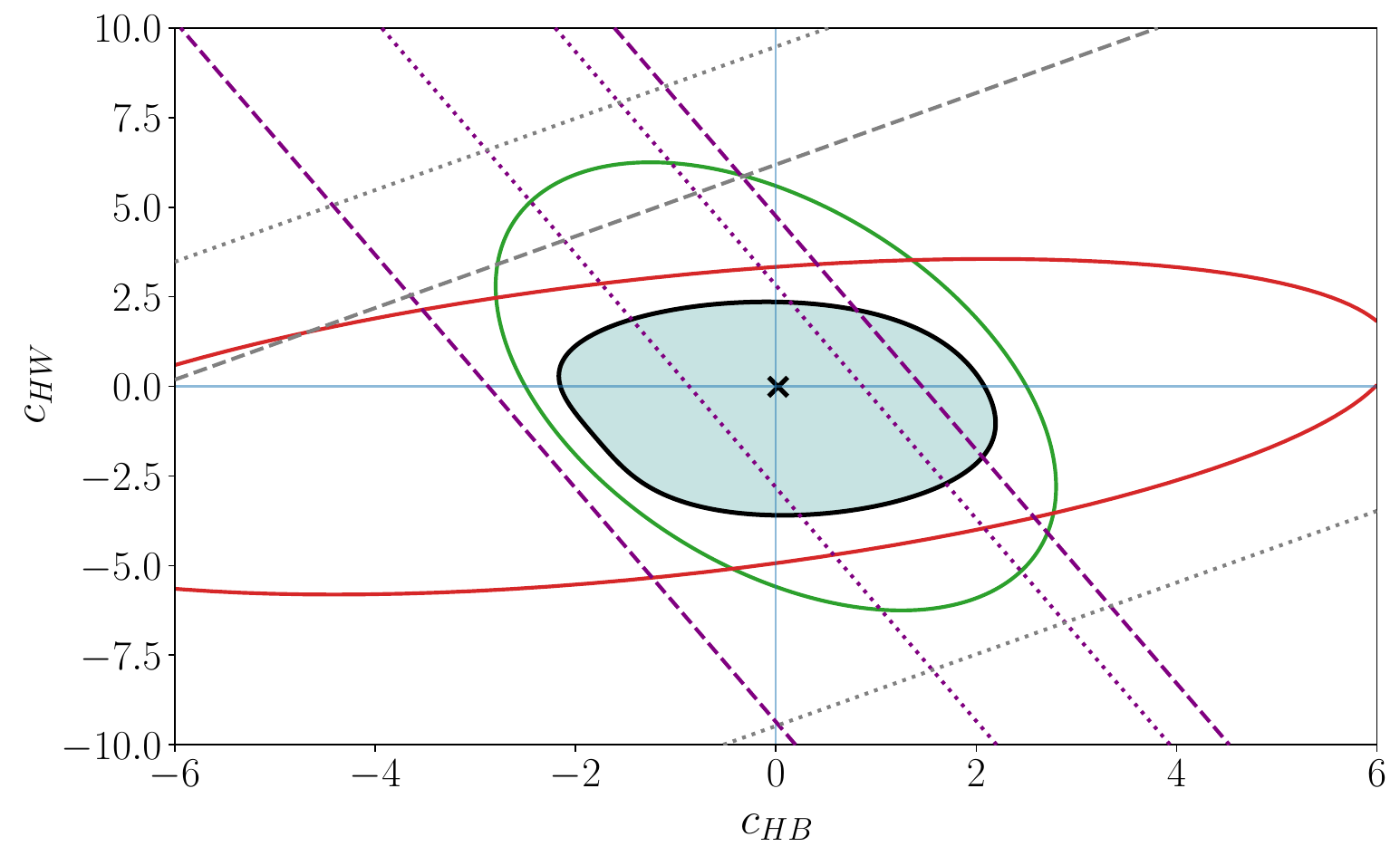}
    \caption{Results of the $\chi^2$ fit for leading triboson processes in the $(c_{HB}, c_{HW})$, $(c_{HB}, c_{HWB})$, and $(c_{HW}, c_{HWB})$ planes. The third SMEFT Wilson coefficient in each panel was set to zero. The filled regions represent the combined $\chi^2$ fit of the analysed triboson processes.} \label{fig:2to3}
\end{figure}
%%%%%%%%%%%%%%%%%%%%%%%%%%%%%%%%%%%%%%%%%%%%%%%%%%%%%%%%%%%%%%%%%%%%%%%%%%%%%%%%%%%%%%%%%%%%%%%%%%%%

In \Cref{fig:2to3} we present the combined $\chi^2$ fit for the leading $2 \to 3$ processes in comparison with the bounds from other probes. The complete results for all processes are given in the Appendix.
We fix $\sqrt{s} = 10\,\mathrm{TeV} $ and the NP scale to the reference value $\Lambda = 20\,\mathrm{TeV}$ for normalization of the Wilson coefficients. The details of the selection cuts used for each process are given in \Cref{tab:cuts}.
The main driving processes are those containing Higgs bosons.
In particular, we find that the processes with the most constraining power are $\ell^+ \ell^- \to hh\gamma $ and $\ell^+ \ell^- \to hhZ$; for the Wilson coefficient $c_{HWB}$ the process $\ell^+ \ell^- \to W^+ W^- h$ is also a relevant contribution constraint.
Notably, the combination of our three-body final state processes does not have significant flat directions, having a similar sensitivity to all three Wilson coefficients.
Comparing to the constraints from rare Higgs decays, this entails that a high energy lepton collider can bring significant advances on the knowledge of the $Z\gamma$ coupling, for which precision measurements have limited sensitivity,
while the $\gamma\gamma$ coupling can only marginally be improved compared to HL-LHC.

%%%%%%%%%%%%%%%%%%%%%%%%%%%%%%%%%%%%%%%%%%%%%%%%%%%%%%%%%%%%%%%%%%%%%%%%%%%%%%%%%%%%%%%%%%%%%%%%%%%%
\begin{table}[t]
\centering
\renewcommand{\arraystretch}{1.3}
\setlength{\tabcolsep}{6pt}
\small
\begin{tabular*}{\textwidth}{
@{\extracolsep{\fill}}
>{$\displaystyle}l<{$}
>{\centering\arraybackslash}m{6.5cm}
>{\centering\arraybackslash}m{4.5cm}
}
\toprule
\textnormal{\textbf{Process}} & \textbf{Invariant mass cut [GeV]} & \textbf{Scattering angle cut} \\
\midrule
\mu^+ \mu^- \to h \gamma
& $m_h > 112$
& $-$ \\
\mu^+ \mu^- \to h Z
& $m_h > 109$
& $-$ \\
\mu^+ \mu^- \to W^+ W^-
& $-$
& $-0.99 < \cos\theta_{W^+} < 0.65$ \\
\mu^+ \mu^- \to Z Z
& $-$
& $-$ \\
\mu^+ \mu^- \to Z \gamma
& $-$
& $-$ \\
\mu^+ \mu^- \to \gamma \gamma
& $-$
& $-$ \\
\mu^+ \mu^- \to h h \gamma
& $m_h > 108$
& $-0.88 < \cos\theta_\gamma < 0.88$ \\
\mu^+ \mu^- \to h \gamma \gamma
& $m_h > 111$ %<140
& $-$ \\
\mu^+ \mu^- \to h h Z
& $m_h > 106$
& $-$ \\
\mu^+ \mu^- \to W^+ W^- \gamma
& $-$
& $-0.99 < \cos\theta_{W^+} < 0.56$ \\
\mu^+ \mu^- \to W^+ W^- h
& $m_h > 109$
& $-$ \\
\mu^+ \mu^- \to W^+ W^- Z
& $m_Z < 104$
& $-0.99 < \cos\theta_{W^+} < 0.46$ \\
\mu^+ \mu^- \to Z Z \gamma
& $-$
& $-0.81 < \cos\theta_Z < 0.81$ \\
\mu^+ \mu^- \to Z \gamma \gamma
& $-$
& $-$ \\
\mu^+ \mu^- \to Z Z Z
& $-$
& $-0.87 < \cos\theta_Z < 0.87$ \\
\mu^+ \mu^- \to Z h \gamma
& $m_h>106$
& $-$ \\
\bottomrule
\end{tabular*}
\caption{Summary of selection cuts applied to diboson and triboson processes.
Here $m_i$ is the invariant mass of particle $i$ after the Gaussian smearing, and $\theta_i$ is its polar scattering angle.}
\label{tab:cuts}
\end{table}
%%%%%%%%%%%%%%%%%%%%%%%%%%%%%%%%%%%%%%%%%%%%%%%%%%%%%%%%%%%%%%%%%%%%%%%%%%%%%%%%%%%%%%%%%%%%%%%%%%%%

Finally, we comment on the unitarity bounds for the operators of interest.
Notice that the $E_{\rm cm}^2/\Lambda^4$ factor of the NP cross-section also helps with the validity of our EFT analysis. Indeed, the scaling is such that the limit on the EFT scale $\Lambda_{95}$ increases linearly with the center-of-mass energy, allowing it to avoid the unitarity bounds at arbitrarily high energies.
The limits from $\ell^+\ell^-\to hh\gamma$ (green) and $\ell^+\ell^-\to hhZ$ (red) are plotted in \Cref{fig:twototwo-and-three-eft-validity} along with the ones from dibosons. The complete set of results for all analysed processes is given in \Cref{fig:app:twototwo-eft,fig:app:twotothree-eft} in the Appendix.
One can see that the triboson contours run parallel to the blue iso-lines of constant UV coupling strength. This means that one can in principle reach sensitivity to indefinitely large NP scales using these triboson measurements by increasing the collider energy, without entering a strong-coupling regime where the EFT breaks down and direct production of new particles has to be considered. Once it is established that a muon collider can be sensitive to scenarios with a given value of $g_{\rm NP}$, this conclusion remains unchanged going to higher energies. Key to this finding is that the luminosity of the collider can grow as per our scaling in \cref{eq:lumi-scaling}.

Eventually, going to very high energies, electroweak radiation will play an important role, and blur the separation between the multi-boson processes considered here and higher multiplicity final states. Nevertheless, appealing to concepts similar to those of QCD jets, and concentrating on hard objects in the final states, we expect that high-energy three-body scatterings will continue to play a most significant role in setting the bounds on the operators in \cref{eq:opHVVSMEFT}.

%%%%%%%%%%%%%%%%%%%%%%%%%%%%%%%%%%%%%%%%%%%%%%%%%%%%%%%%%%%%%%%%%%%%%%%%%%%%%%%%%%%%%%%%%%%%%%%%%%%%
\subsection{Combined results}
\label{sec:results}

\begin{figure}[t]
    \centering
    \includegraphics[width=0.5\linewidth]{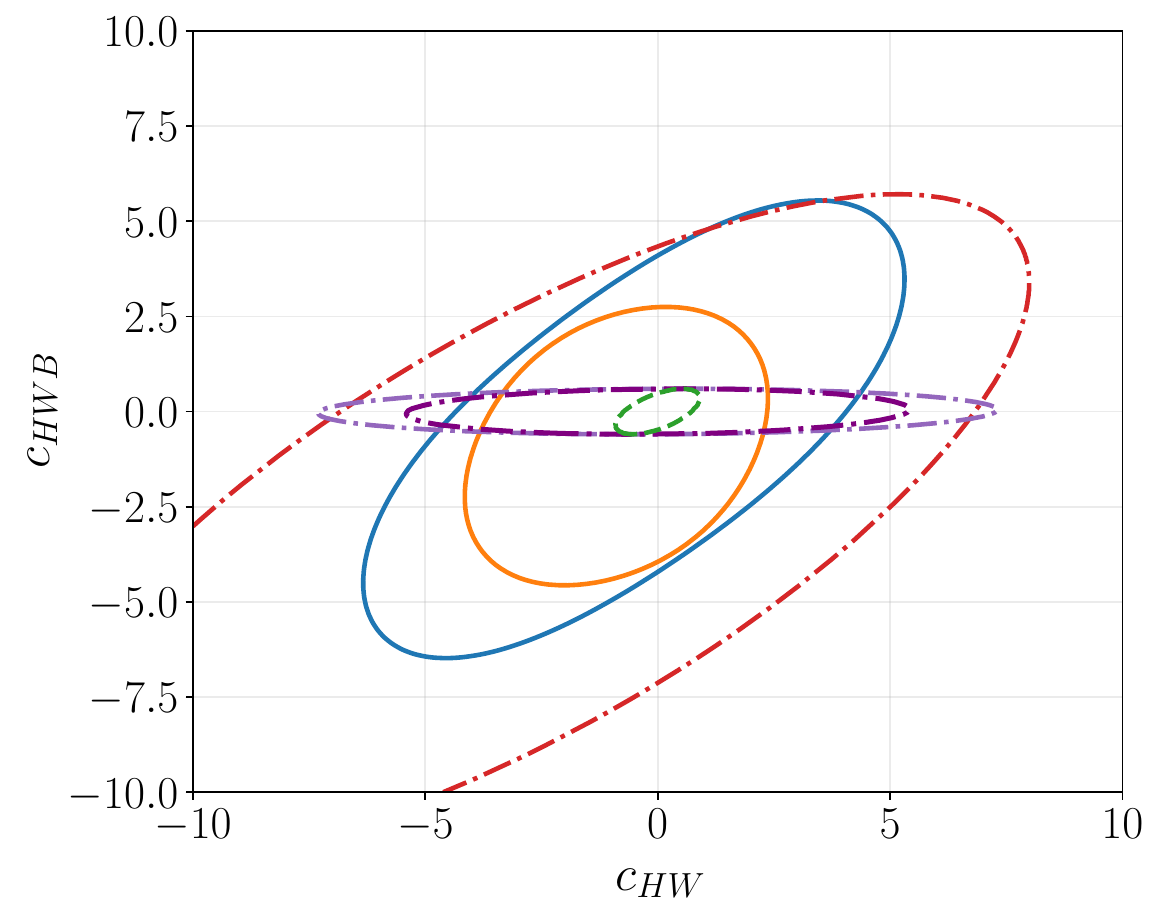}\hfill%
    \includegraphics[width=0.5\linewidth]{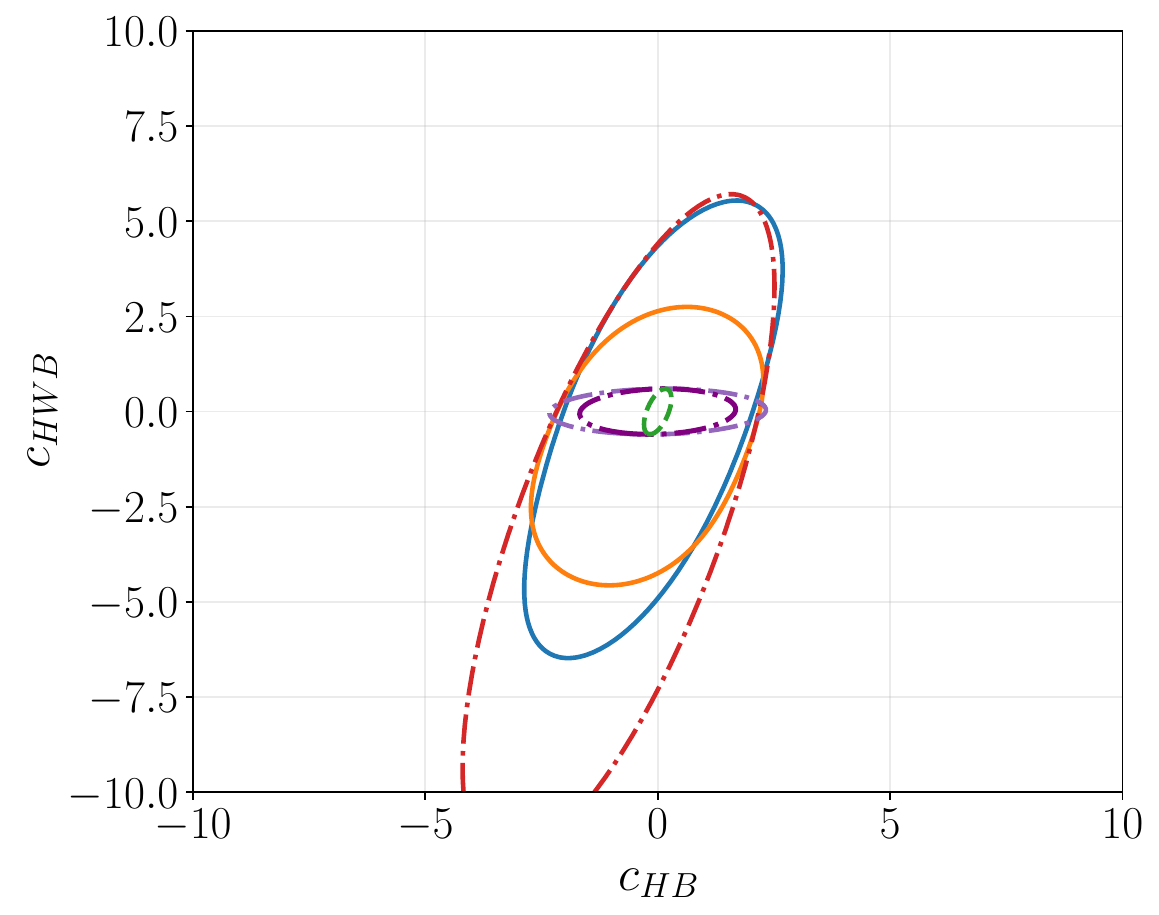}\\
    \raisebox{0.1\height}{\includegraphics[width=0.45\linewidth]{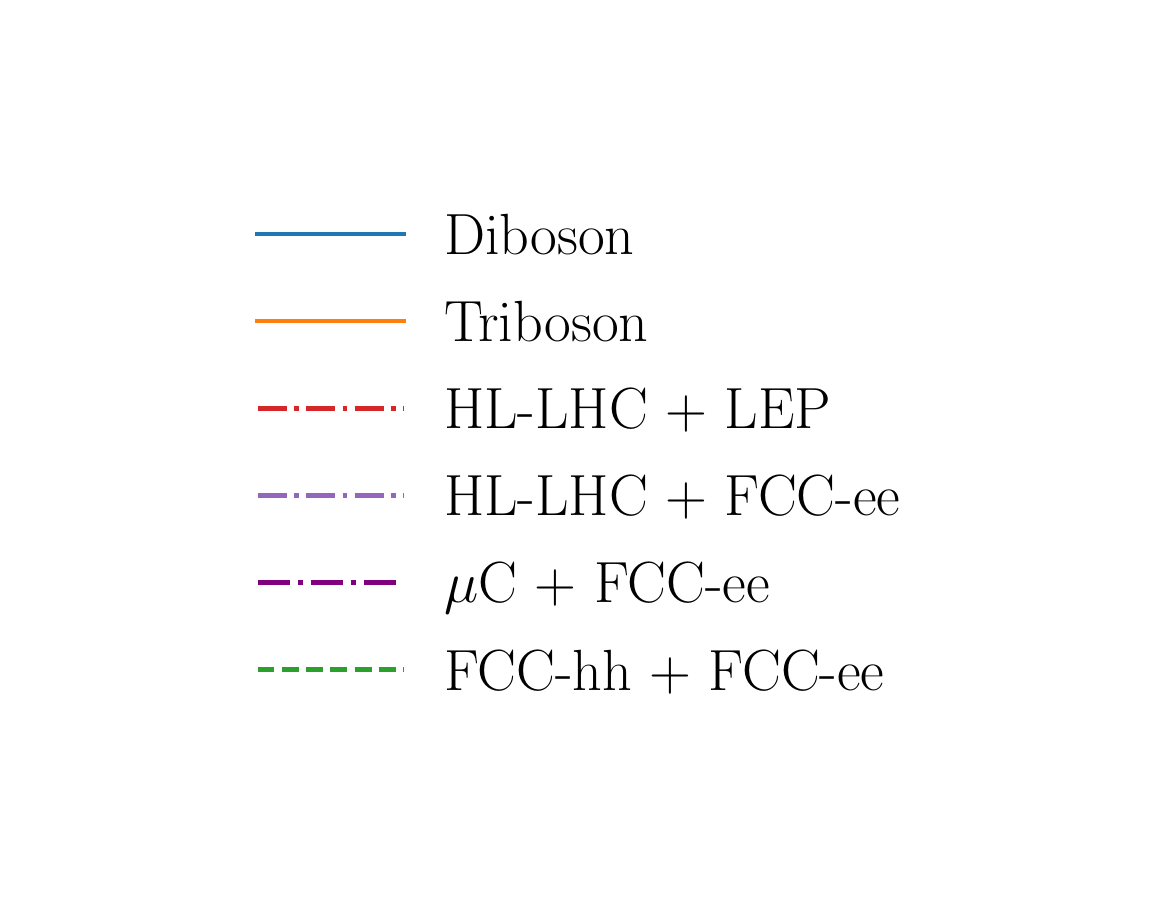}}\hfill%
    \includegraphics[width=0.5\linewidth]{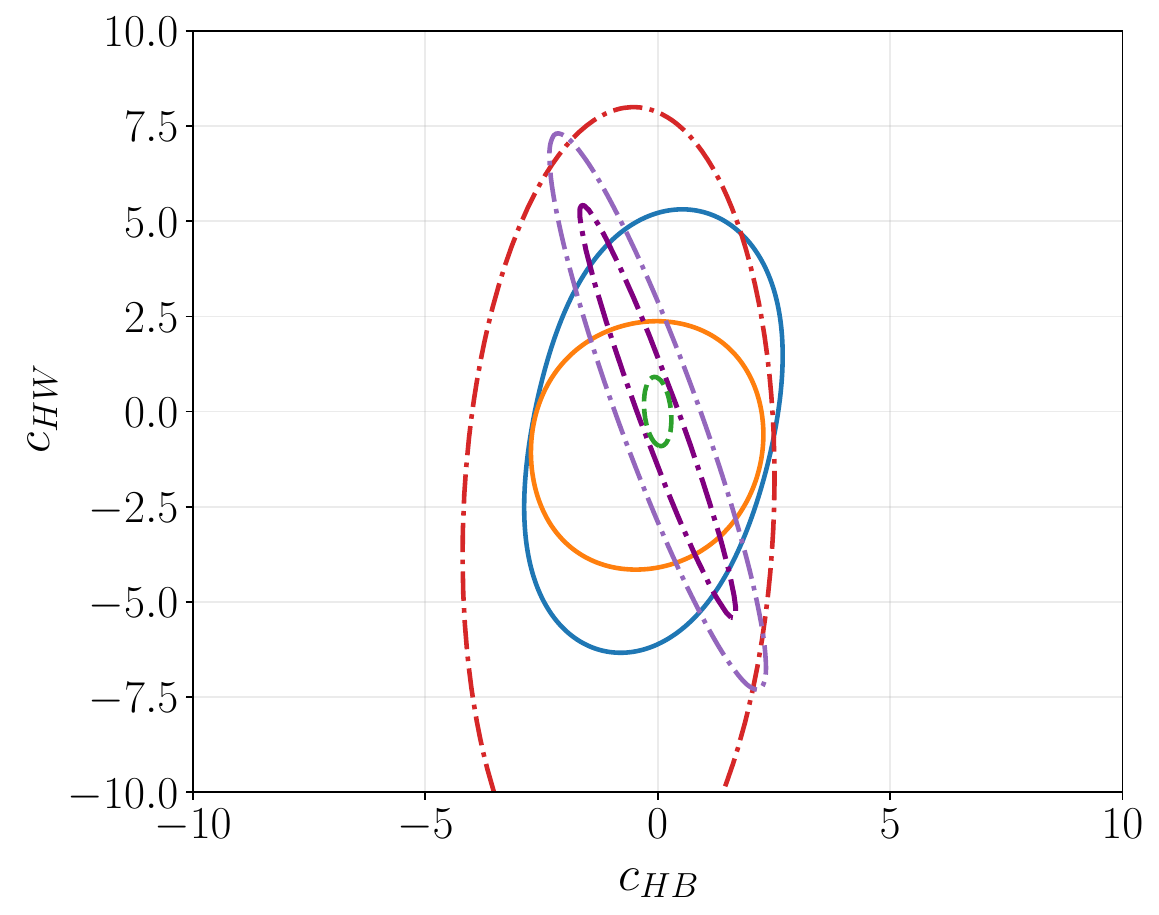}
    \caption{Combined $\chi^2$ fit for all diboson (blue ellipse) and triboson processes (orange ellipse) at $\sqrt{s} = 10$~TeV. The precision constraints from Higgs couplings combined with EWPO are shown as dot-dashed lines: current LEP bounds~\cite{Brivio:2017bnu} combined with HL-LHC~\cite{deBlas:2025gyz} (red), and projections for EWPO at FCC-ee~\cite{Celada:2024mcf} combined with Higgs couplings at HL-LHC (lilac), at a 10~TeV muon collider~\cite{deBlas:2022ofj} (purple), and at FCC-hh~\cite{deBlas:2025gyz,FCC:2018byv, deBlas:2019rxi} (green). In each panel we marginalize over the remaining SMEFT Wilson coefficient.}
    \label{fig:2dprofiled}
\end{figure}

We now compare the sensitivity of diboson and triboson processes with the projected precision constraints from the rare Higgs decays $h \to \gamma\gamma$ and $h \to Z\gamma$ at HL-LHC. \Cref{fig:2dprofiled} shows a comparison of the resulting bounds from diboson processes (blue), triboson processes (orange), and a combination of precision constraints.  

As discussed above, the two Higgs decays alone do not constrain the three couplings of interest in \cref{eq:opHVVSMEFT}. Therefore we combine them with current EWPO bounds obtained from LEP~\cite{Brivio:2017bnu} (red), or projected EWPO bounds from FCC-ee \cite{Celada:2024mcf} (lilac). 
Additionally, we also display the projections obtained in \cite{deBlas:2022ofj} for the rare Higgs decays at a muon collider (purple), where the Higgs bosons are produced in vector boson fusion, with expected precision of $\delta\kappa_\gamma \sim 0.6\%$ and $\delta\kappa_{Z\gamma} \sim 5.2\%$ on the modified couplings.

We show the constraints in the three possible two-dimensional planes, marginalizing over the remaining SMEFT Wilson coefficient. As a consequence of this marginalization, EWPO play a crucial role among precision observables. Present data yield the very loose constraints shown by the red contours, which has to be compared to the results for fixed Wilson coefficients shown in \Cref{fig:2to2,fig:2to3}. This will change with the extreme precision expected at FCC-ee on EWPO, which effectively will fix $c_{HWB}$ to be close to zero. The purple and lilac contours in \Cref{fig:2dprofiled} are thus very similar to the Higgs couplings results of \Cref{fig:2to2,fig:2to3}, while the red contours are much wider.

%%%%%%%%%%%%%%%%%%%%%%%%%%%%%%%%%%%%%%%%%%%%%%%%%%%%%%%%%%%%%%%%%%%%%%%%%%%%%%%%%%%%%%%%%%%%%%%%%%%%
%%%%%%%%%%%%%%%%%%%%%%%%%%%%%%%%%%%%%%%%%%%%%%%%%%%%%%%%%%%%%%%%%%%%%%%%%%%%%%%%%%%%%%%%%%%%%%%%%%%%

In this picture, our results from multi-boson production, in particular tribosons, nicely complement those from EW and Higgs precision measurements at low energies.
Indeed, in all the plots the orange contour is narrower than the purple or lilac contours along some direction. This means that by combining the processes studied here with Higgs measurements one can significantly reduce the allowed SMEFT parameter space. This remains true even if one includes the improvements in Higgs couplings possible at the muon collider itself.
Notice however that, due to its huge Higgs production rate, a future very-high-energy hadron collider has the potential of significantly surpassing all these constraints by measuring the rare Higgs decays alone. The green contours in \Cref{fig:2dprofiled} show the expected reach of FCC-hh at $\sqrt{s} = 85$~TeV.
It is worth mentioning that high-energy diboson and triboson scatterings at the same FCC-hh could still play an important and complementary role. We leave a precise analysis of this point to a future study.

\begin{figure}[t]
\centering%
\includegraphics[width = 0.5\linewidth]{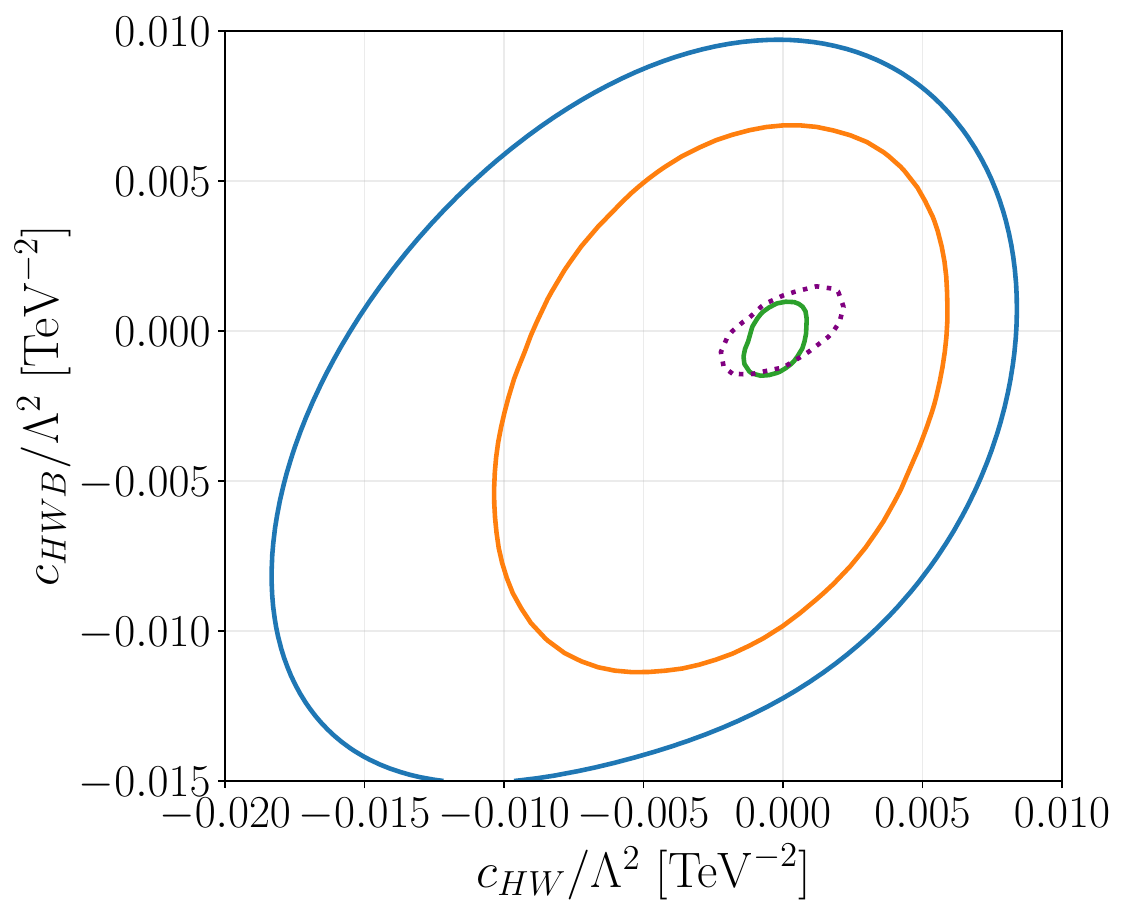}\hfill%
\includegraphics[width = 0.5\linewidth]{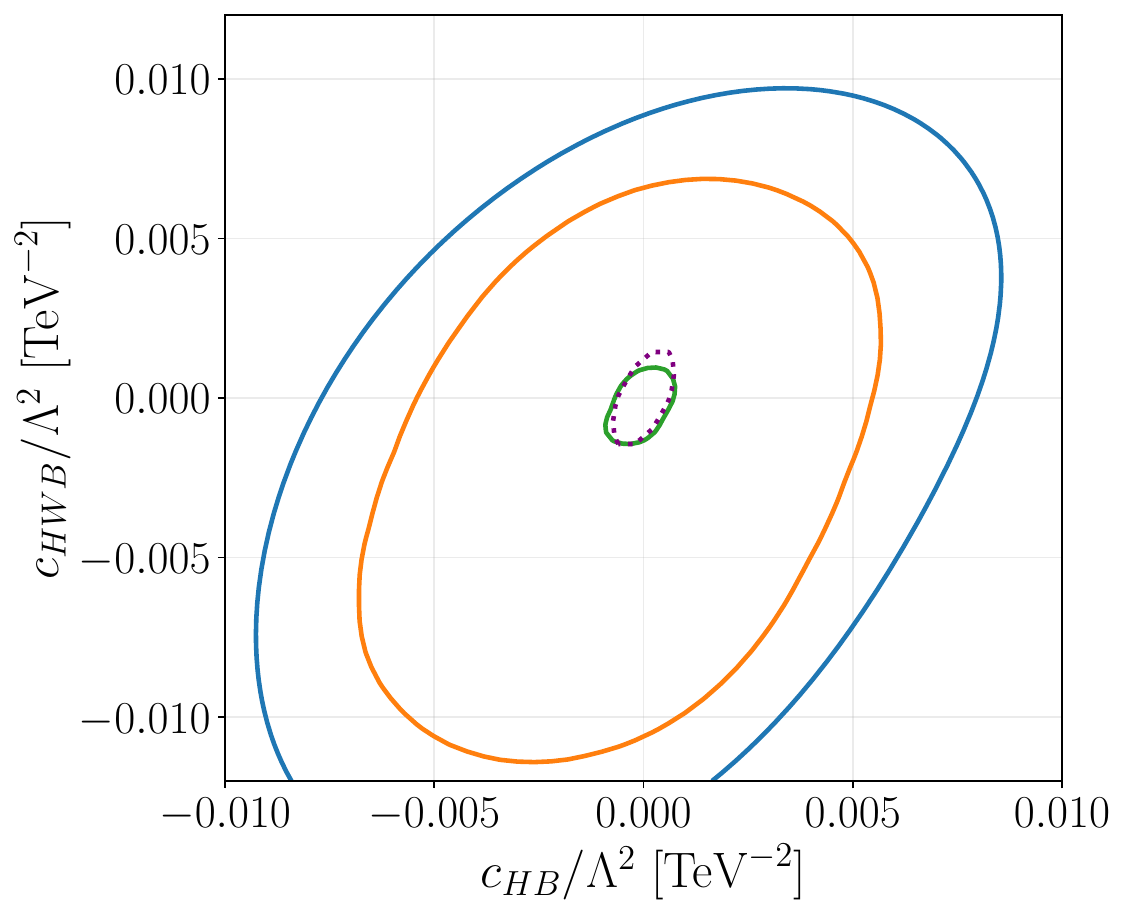}\\
\raisebox{0.1\height}{\includegraphics[width=0.45\linewidth]{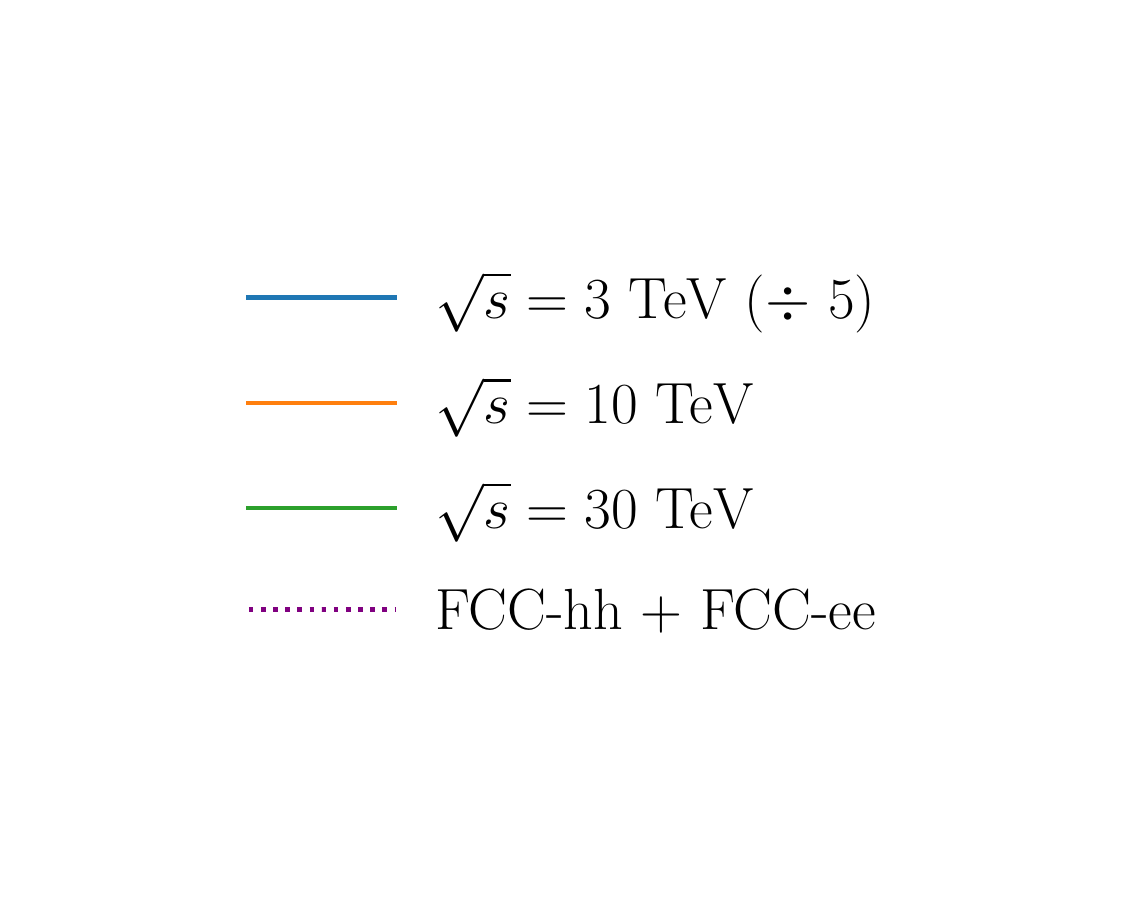}}\hfill%
\includegraphics[width=0.5\linewidth]{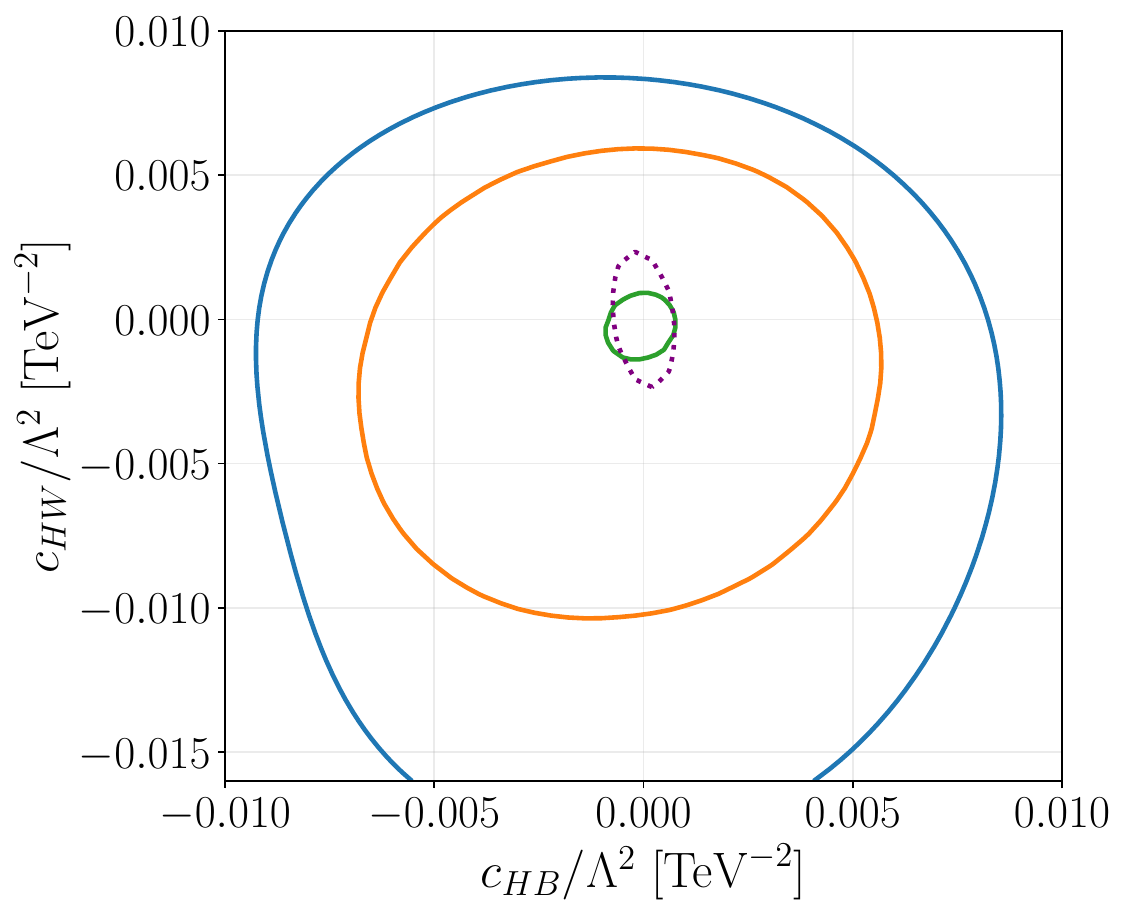}
\caption{Results of the total $\chi^2$ fit for triboson processes at a muon collider with $3$~TeV (blue), $10$~TeV (orange), and $30$~TeV (green) center-of-mass energy, and for FCC-hh+FCC-ee (dotted). The bounds are shown in the $(c_{HB}, c_{HW})$, $(c_{HB}, c_{HWB})$, and $(c_{HW}, c_{HWB})$ planes, marginalising over the remaining Wilson coefficient. The 3~TeV bounds are scaled by a factor 5 on both axes.}
    \label{fig:3to2_fcchhee_marg}
\end{figure}

In addition to the complementarity discussed above, we reiterate that triboson production retains its sensitivity at higher collider energies without violating the EFT constraints.
In \Cref{fig:3to2_fcchhee_marg} we show the scaling of the expected bounds on the Wilson coefficients for different values of the muon collider center-of-mass energy, along with the known projections for FCC-hh from Higgs rare decays at 85 TeV~\cite{deBlas:2025gyz,FCC:2018byv,deBlas:2019rxi}. The FCC-hh projections suggest a sub-percent precision on both couplings, namely $\delta \kappa_\gamma \sim 0.3\%$ and $\delta \kappa_{Z\gamma} \sim 0.7 \%$.
We find that a lepton collider running at $\sqrt{s} = 30$ TeV can be comparable, or even slightly more powerful, in constraining the parameter space of our interest than FCC-hh.

%%%%%%%%%%%%%%%%%%%%%%%%%%%%%%%%%%%%%%%%%%%%%%%%%%%%%%%%%%%%%%%%%%%%%%%%%%%%%%%%%%%%%%%%%%%%%%%%%%%%
%%%%%%%%%%%%%%%%%%%%%%%%%%%%%%%%%%%%%%%%%%%%%%%%%%%%%%%%%%%%%%%%%%%%%%%%%%%%%%%%%%%%%%%%%%%%%%%%%%%%

%%%%%%%%%%%%%%%%%%%%%%%%%%%%%%%%%%%%%%%%%%%%%%%%%%%%%%%%%%%%%%%%%%%%%%%%%%%%%%%%%%%%%%%%%%%%%%%%%%%%
\section{Conclusions}
\label{sec:conclusions}
%%%%%%%%%%%%%%%%%%%%%%%%%%%%%%%%%%%%%%%%%%%%%%%%%%%%%%%%%%%%%%%%%%%%%%%%%%%%%%%%%%%%%%%%%%%%%%%%%%%%

In this work, we have analyzed a set of processes involving the Higgs boson and Standard Model gauge bosons in the final state at a future high-energy lepton collider. Our goal was to assess the sensitivity of such processes to  EFT Higgs interactions arising from new physics, focusing on a subset of dimension-six operators in the SMEFT Warsaw basis. The choice of operators has been driven by the intention to test new physics hypotheses that give rise to modifications in rare Higgs boson decays, in particular those originated at loop level in the SM, $h\to\gamma\gamma$ and $h \to Z\gamma$. This restricted our interest to the three SMEFT operators given in \cref{eq:opHVVSMEFT}. 

These operators have the schematic form $|H|^2 F_{\mu\nu}F^{\mu\nu}$ and can therefore arise naturally in perturbative extensions of the SM through loop diagrams involving new electroweak states coupled to the Higgs sector.
If such new particles have a symmetry that forbids them from coupling linearly to the SM, the interactions in \cref{eq:opHVVSMEFT} arise very naturally with a size similar to the one of other operators. In concrete scenarios, they can e.g.\ be originated from loops of states that belong to a dark sector, %comprising Dark Matter particles,
or in supersymmetric models in which linear couplings of superpartners to SM particles is forbidden by R-parity.
While in these scenarios the relevant operators always arise at one loop, one can easily envisage situations where a strong coupling or a large number of states counteracts the loop suppression, yielding the estimate
\begin{equation}
    \Lambda \sim M_{\rm NP}/g_{\rm NP}
\end{equation}
with $g_{\rm NP}$ a coupling of order one, and $M_{\rm NP}$ the typical mass scale of the new physics.

We derived bounds on the three operators \cref{eq:opHVVSMEFT} from a wide range of multi-boson production processes at a future multi-TeV lepton collider. Different processes can exhibit distinct behaviours as the collider energy is varied, as illustrated in \Cref{fig:twototwo-and-three-eft-validity}.
The projected bounds are shown in the $(\sqrt{s},\Lambda)$ plane together with contours of constant UV coupling strength, allowing us to track their energy dependence and assess their relation to the region of perturbative EFT validity.
We found that triboson processes have the notable property to give bounds corresponding to the same UV coupling strength $g_{UV}$ at all center-of-mass energies, provided the luminosity follows \cref{eq:lumi-scaling}, thus allowing to improve the constraints indefinitely by increasing the collider energy. Among these, production of two Higgs boson and a neutral gauge boson $\ell^+\ell^- \to Vhh$ are particularly useful due to their lower SM background, and give the strongest constraints at $\sqrt{s} = 10$~TeV.
All other processes, including diboson production $\ell^+\ell^-\to VV,Vh$ can give useful bounds up to collider energies of $10\!-\! 20$~TeV, and become less effective at higher energies, requiring larger couplings to generate NP effects within the collider reach.
In view of these results, we have concluded that the most effective route to probe operators from \cref{eq:opHVVSMEFT} at a high-energy lepton collider is not to use $2\to 2$ scattering processes, as is the case for the operators $\cO_B$ and $\cO_W$ studied in~\cite{Buttazzo:2020uzc}, but rather to use $2\to 3$ processes.

Our results for the three considered Wilson coefficients at $\sqrt{s} = 10$~TeV are shown in \Cref{fig:2to2,fig:2to3} for diboson and triboson processes, respectively. We find that $\ell^+\ell^-\to h \gamma$ provides the strongest constraints among the two-body final states for all three Wilson coefficients. For the three-body final states, the process $\ell^+ \ell^- \to h h \gamma$ emerges as the most sensitive channel, with $\ell^+ \ell^- \to h hZ$ also contributing significantly to the overall constraint. The strong performance of the photon plus Higgs channels can be primarily attributed to their negligible Standard Model backgrounds, which allows for highly efficient signal extraction. A key to our results is the ability to experimentally separate hadronic Higgs and $Z$ boson decays, which relies mostly on di-jet mass resolution.

In \Cref{fig:2dprofiled} we show our combined results marginalized over all the Wilson coefficients. We compared them with constraints from low-energy precision observables, notably rare Higgs decays at the LHC and electroweak precision observables at LEP, and with their expected improvements at future high intensity $e^+e^-$ machines such as FCC-ee.
We find that high-energy lepton colliders offer stronger sensitivity along the directions in EFT parameter space that correspond to the ones probed by the $h \to Z\gamma$ decay. Instead, the diphoton decay channel $h \to \gamma \gamma$ at the HL-LHC is expected to provide the tightest constraints in its corresponding direction of EFT interactions parameter space.
Finally, \Cref{fig:3to2_fcchhee_marg} shows projections for different collider energies.
Exploiting the energy-growing sensitivity of triboson processes, we compare projections for $\sqrt{s}=3,\,10,$ and $30$~TeV muon colliders with the expected FCC-hh sensitivities to the effective couplings $\kappa_{Z\gamma}$ and $\kappa_\gamma$, supplemented by electroweak precision measurements from FCC-ee.
We find that a lepton collider operating at $\sqrt{s} = 30 $~TeV can achieve comparable or even slightly better sensitivity to these couplings than the combined FCC program.

It should be remarked that high-energy colliders also have an advantage in improving the rare Higgs decay measurements of $h\to\gamma\gamma$ and $h\to Z\gamma$, due to the increased rate of Higgs production.
This holds both for high-energy lepton colliders such as the muon collider, where large numbers of Higgs bosons are produced in vector boson fusion, and for hadron colliders. We find that for a 10~TeV muon collider the main improvement comes from the high-energy probes studied here, although with a non-negligible input from Higgs decay observables. A 100~TeV hadron collider, on the other hand, can do much better with Higgs decays alone due to its enormous Higgs production rate~\cite{deBlas:2025gyz,FCC:2018byv,deBlas:2019rxi}.
Notice that Higgs decays always probe the EFT interactions at a low scale around $m_h$, independently of the energy of the collider, thus the strength of the constraint is strictly correlated with the precision of the measurement.
Therefore, even for rare processes as the ones under consideration here, there is an intrinsic limit given by systematic uncertainties on the possible improvement of precision measurements at low $p_{\rm T}$. On the contrary, high-$p_{\rm T}$ scatterings such as the triboson processes (and dibosons, with the limitations discussed above) keep improving the information gain on NP as long as the collider energy is increased, without relying on extreme experimental precisions.

Overall, our results highlight the potential of high-energy lepton colliders as powerful probes of new physics in the Higgs sector. In particular, multi-boson final states provide a complementary handle on NP compared to Higgs couplings measurements, as in general they probe different directions in the SMEFT parameter space. In addition, triboson processes can give bounds that do not saturate perturbative unitarity as the collider center-of-mass energies increase, thus extending the reach to larger and larger scales. The inclusion of electroweak corrections eventually may affect our findings, although the effects are expected to be mild for energies of order 10~TeV. We leave the study of these effects for future works.

%%%%%%%%%%%%%%%%%%%%%%%%%%%%%%%%%%%%%%%%%%%%%%%%%%%%%%%%%%%%%%%%%%%%%%%%%%%%%%%%%%%%%%%%%%%%%%%%%%%%
\section*{Acknowledgements}
This work is supported in part by the European Union - Next
Generation EU and MUR
%Missione 4 - Componente 2 - Investimento 1.1  CUP: I53D23000950006
through the PRIN2022 grant n.202289JEW4.

%%%%%%%%%%%%%%%%%%%%%%%%%%%%%%%%%%%%%%%%%%%%%%%%%%%%%%%%%%%%%%%%%%%%%%%%%%%%%%%%%%%%%%%%%%%%%%%%%%%%
\appendix
%%%%%%%%%%%%%%%%%%%%%%%%%%%%%%%%%%%%%%%%%%%%%%%%%%%%%%%%%%%%%%%%%%%%%%%%%%%%%%%%%%%%%%%%%%%%%%%%%%%%

%%%%%%%%%%%%%%%%%%%%%%%%%%%%%%%%%%%%%%%%%%%%%%%%%%%%%%%%%%%%%%%%%%%%%%%%%%%%%%%%%%%%%%%%%%%%%%%%%%%%
\section{Complete list of bounds}
%\section{EFT validity for diboson and triboson processes}
\label{app:EFTvalidity}

\begin{figure}[h!]
    \centering%
\includegraphics[width=0.48\textwidth]{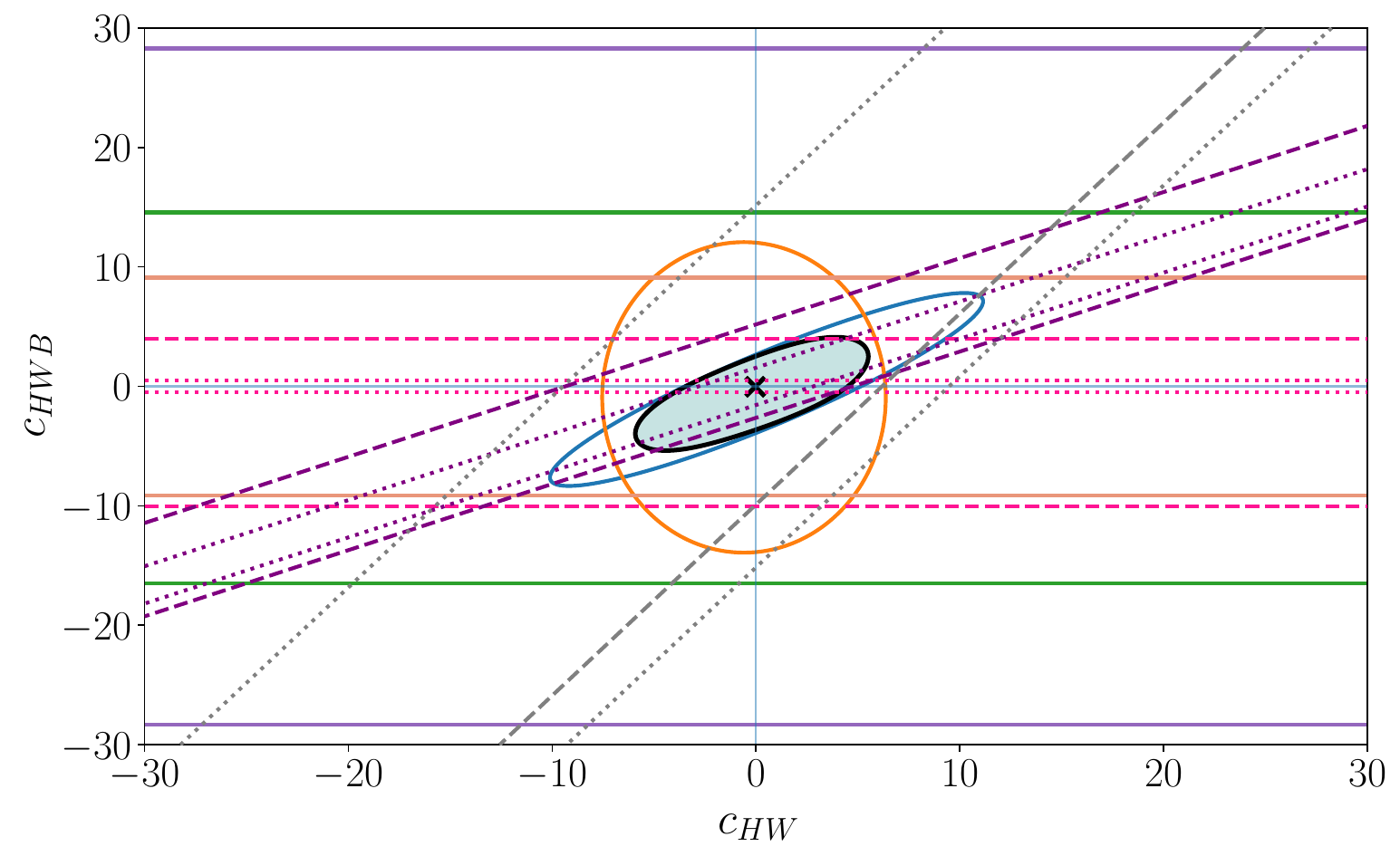}\hfill%
\includegraphics[width=0.48\textwidth]{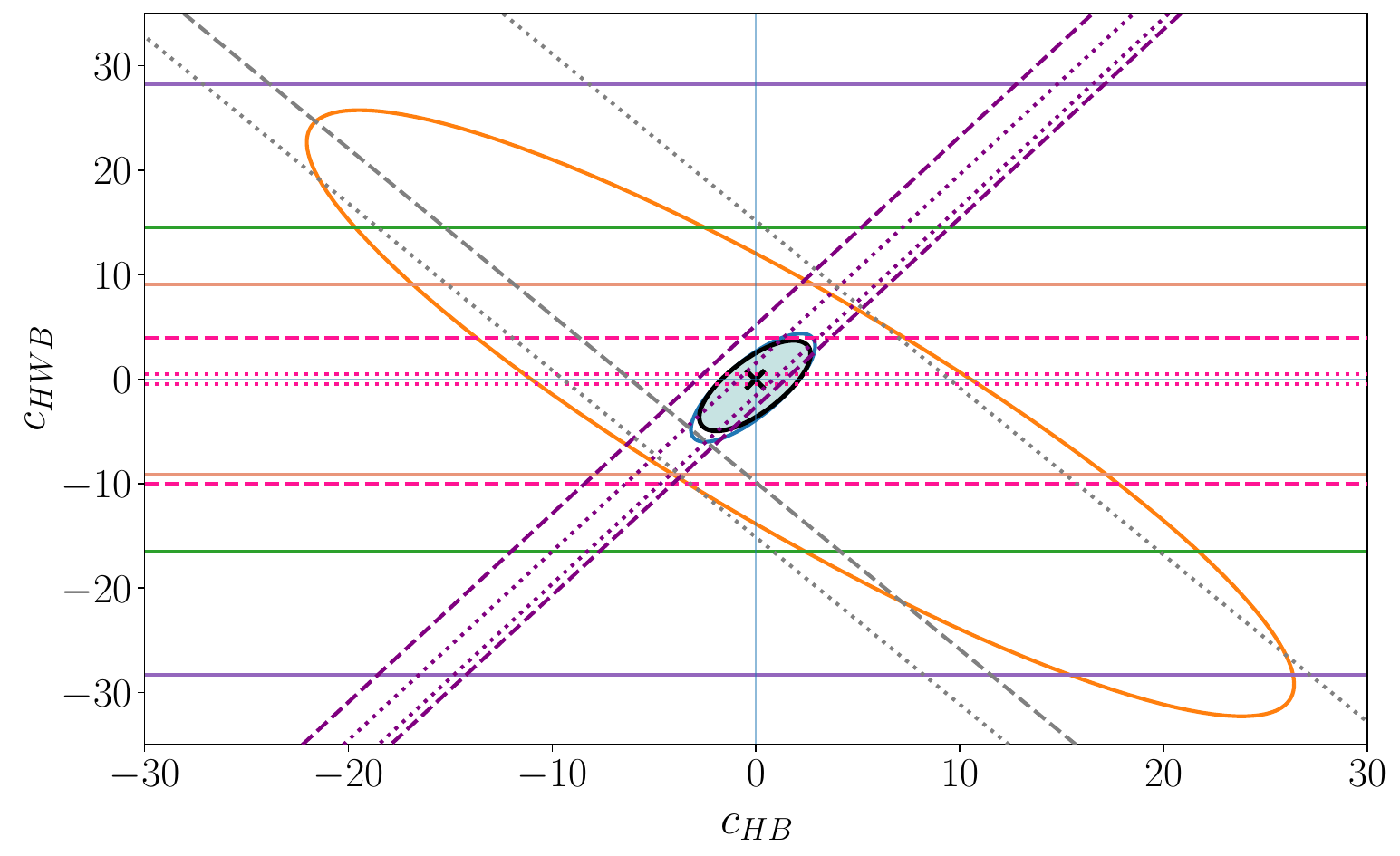}\\
\includegraphics[width=0.48\textwidth]{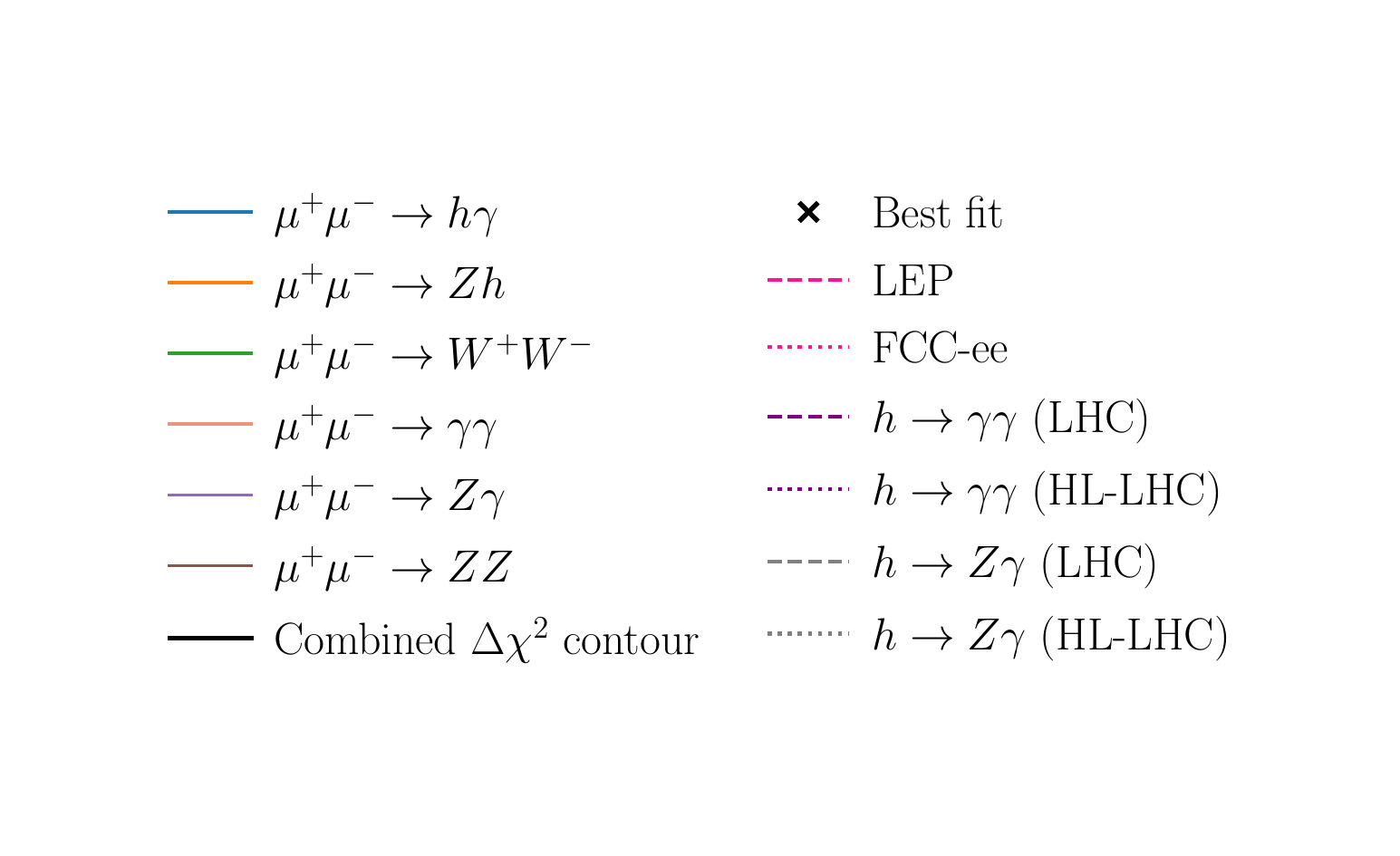}\hfill
\includegraphics[width=0.48\textwidth]{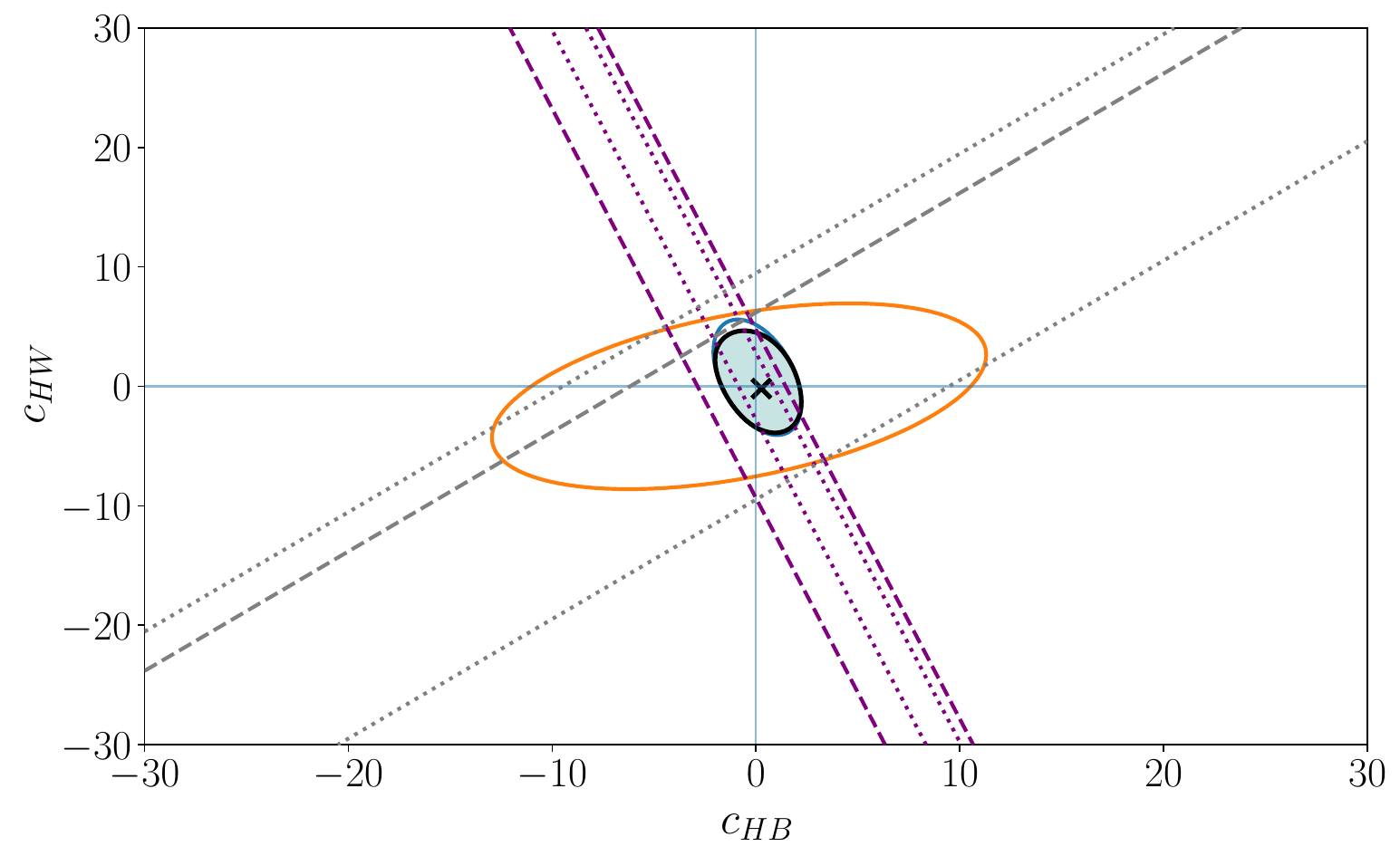}
\caption{Results of the $\chi^2$ fit for dibosons in the $(c_{HB}, c_{HW})$, $(c_{HB}, c_{HWB})$, and $(c_{HW}, c_{HWB})$ planes. The third Wilson coefficient was set to zero in each panel. The filled regions represent the combined fit for $\mu^+ \mu^- \to VV$ and $\mu^+ \mu^- \to Vh$.}
\label{fig:app:2to2total}
\end{figure}

\begin{figure}[h!]
    \centering
\includegraphics[width=0.48\textwidth]{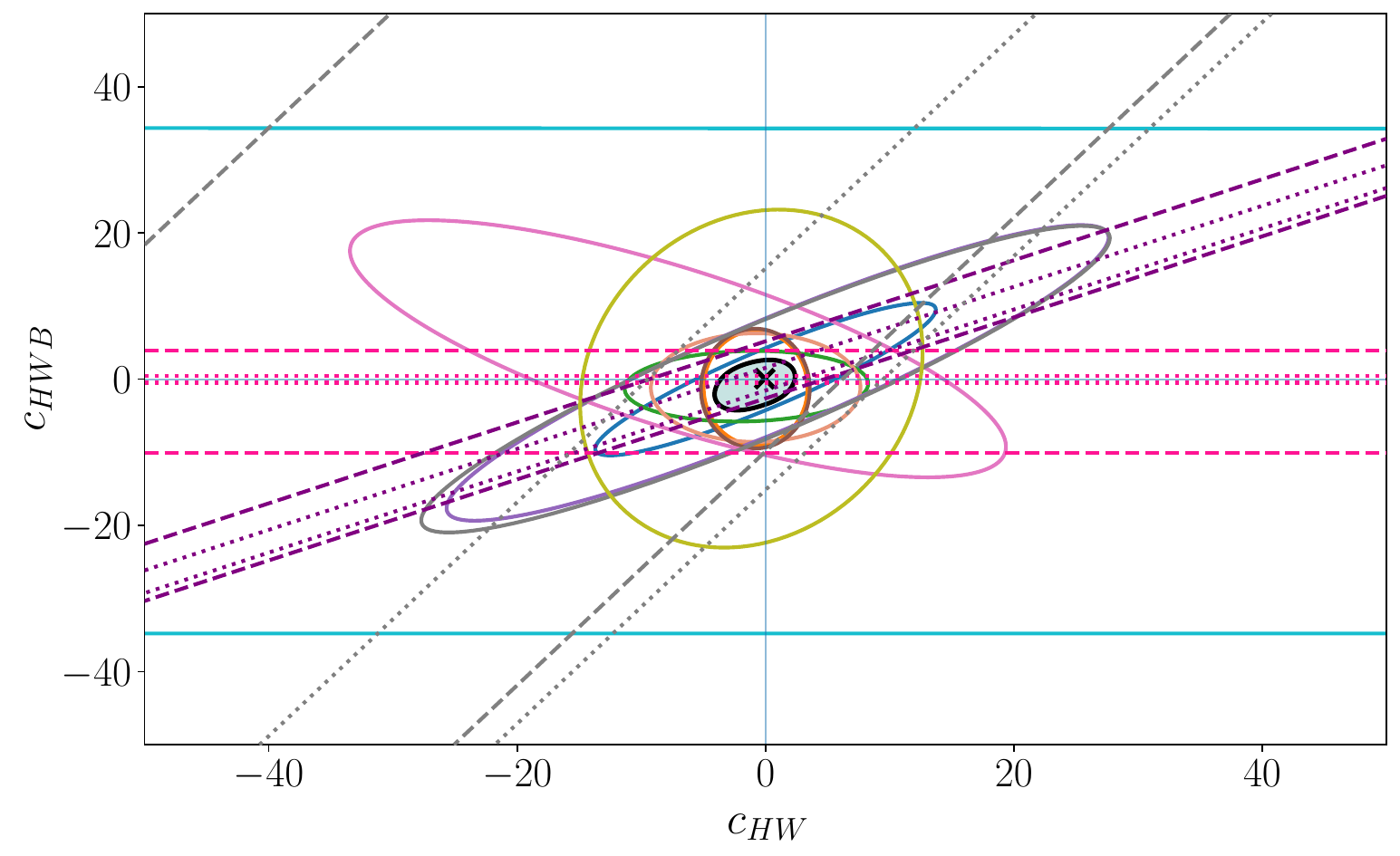}\hfill%
\includegraphics[width=0.48\textwidth]{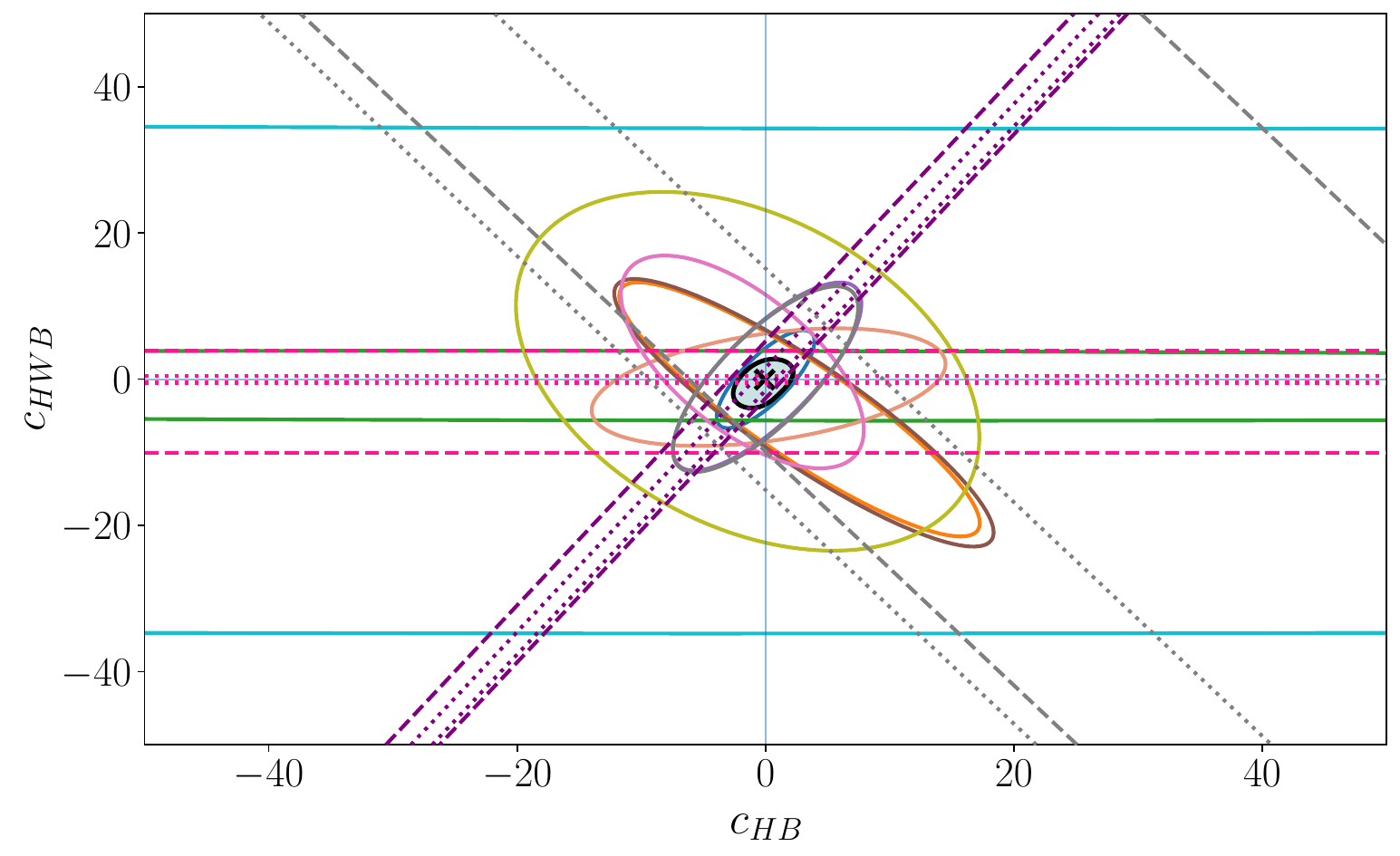}\\
\includegraphics[width=0.48\textwidth]{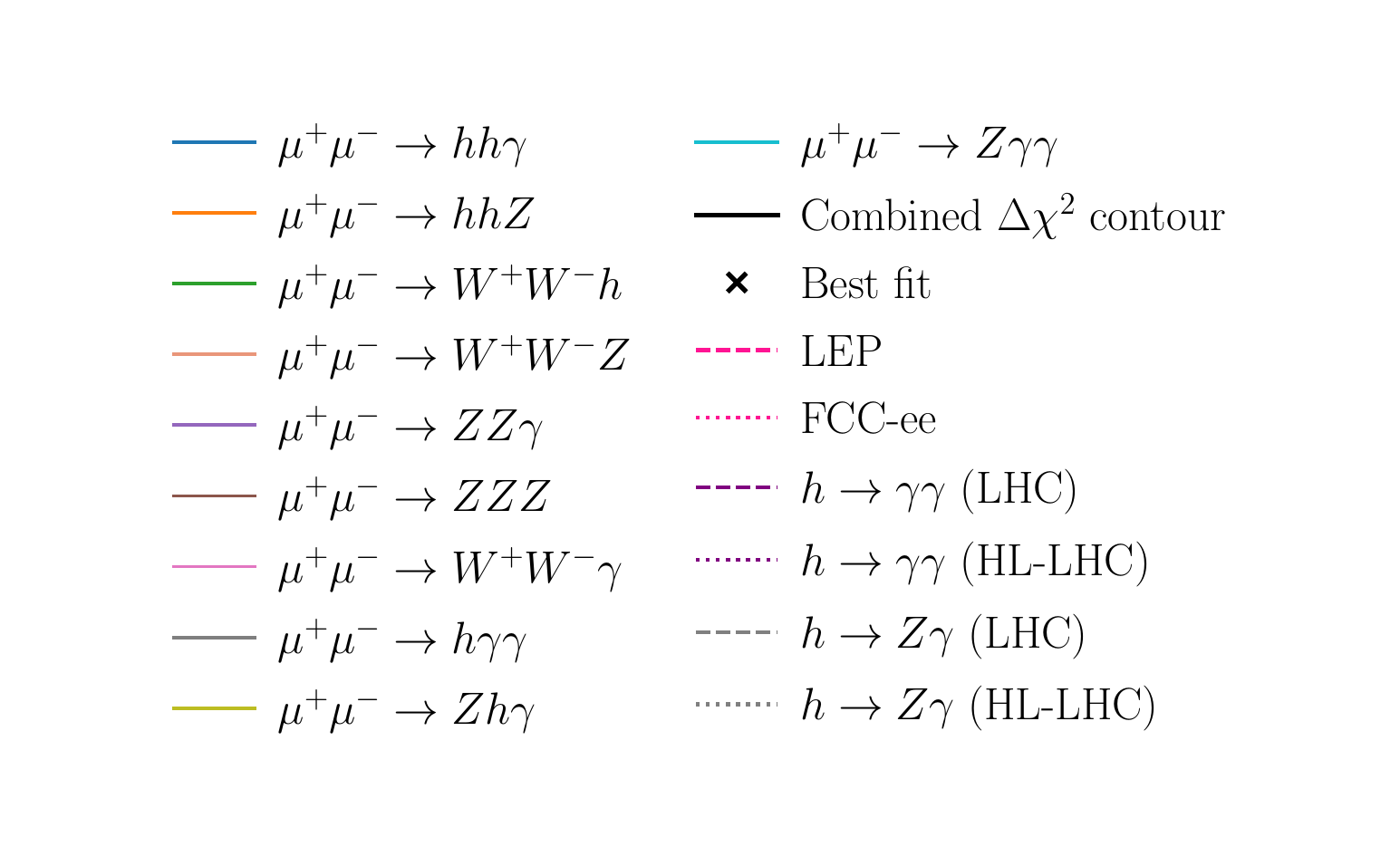}\hfill
\includegraphics[width=0.48\textwidth]{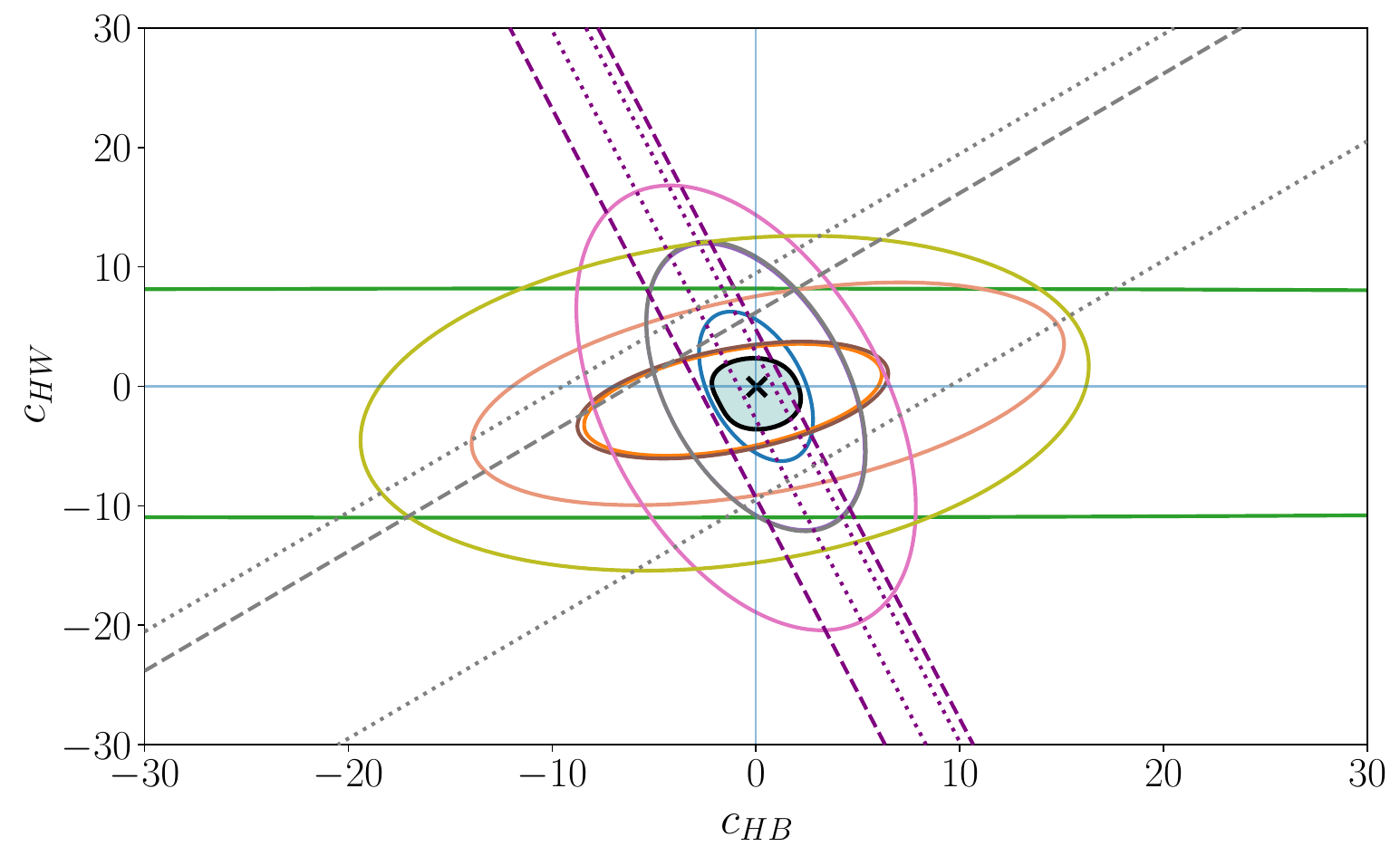}
    \caption{Results of the $\chi^2$ fit for tribosons in the $(c_{HB}, c_{HW})$, $(c_{HB}, c_{HWB})$, and $(c_{HW}, c_{HWB})$ planes. The third Wilson coefficient was set to zero in each panel. The filled regions represent the combined fit for all analysed triboson processes.}
    \label{fig:app:2to3total}
\end{figure}

In this Appendix we provide the bounds for all processes and all operators considered in this paper.
In \Cref{fig:app:2to2total} and \Cref{fig:app:2to3total} below we present the bounds on the Wilson coefficients from all analysed diboson and triboson processes in the same two-dimensional planes as in \Cref{fig:2to2,fig:2to3}.

\begin{figure}[h!]
    \centering%
    \includegraphics[width=0.8\textwidth]{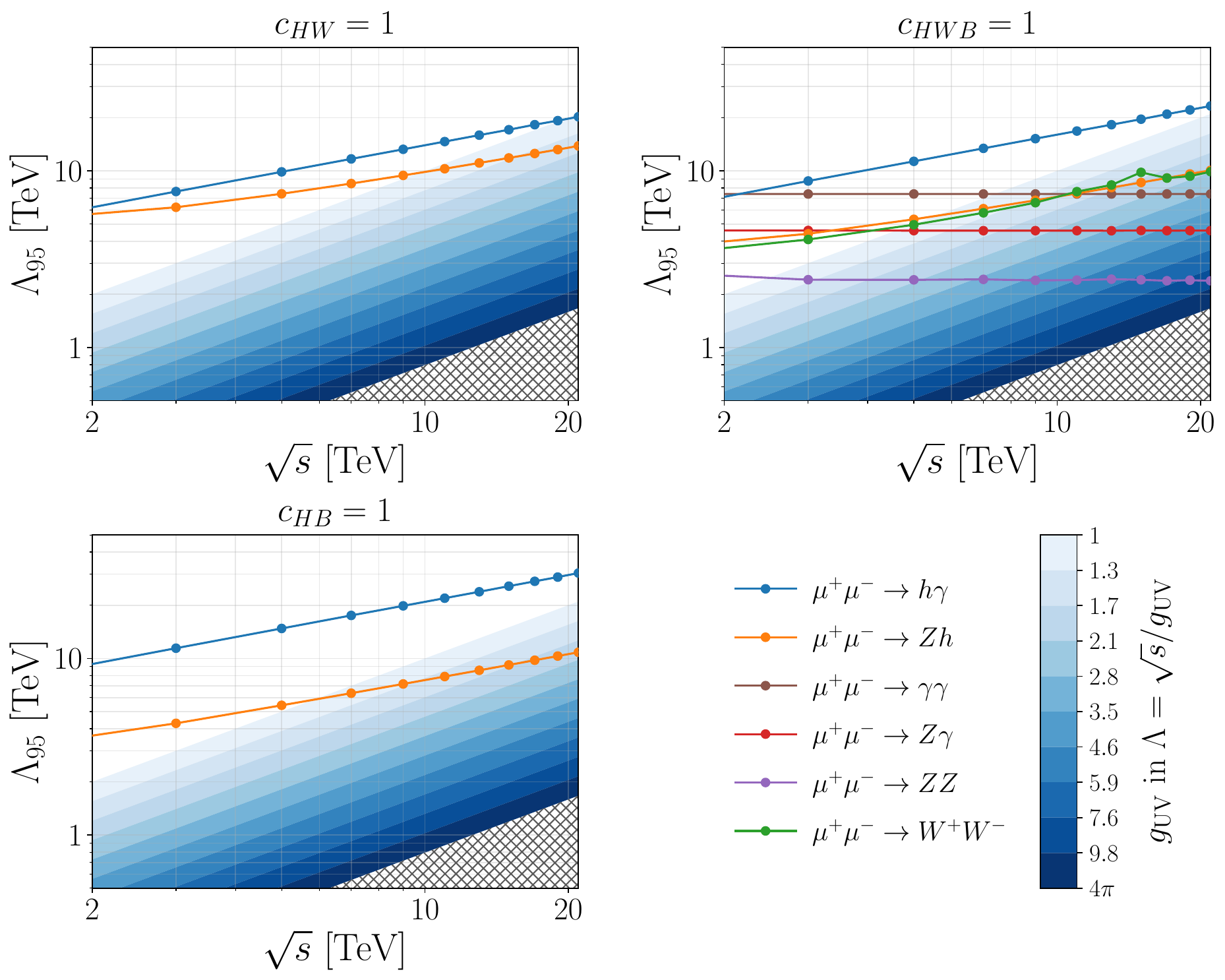}
    \caption{95\% CL lower bound on the EFT scale $\Lambda_{95}$ as a function of $\sqrt{s}$, for all diboson
    processes. In each panel, a single operator with $c_i = 1$ is considered, and the remaining two Wilson coefficients are set to zero. Isolines of constant $g_{\rm UV} = \sqrt{s}/\Lambda$ are shown in blue.
    \label{fig:app:twototwo-eft}}
\end{figure}

\begin{figure}[h!]
    \centering%
    \includegraphics[width=0.8\textwidth]{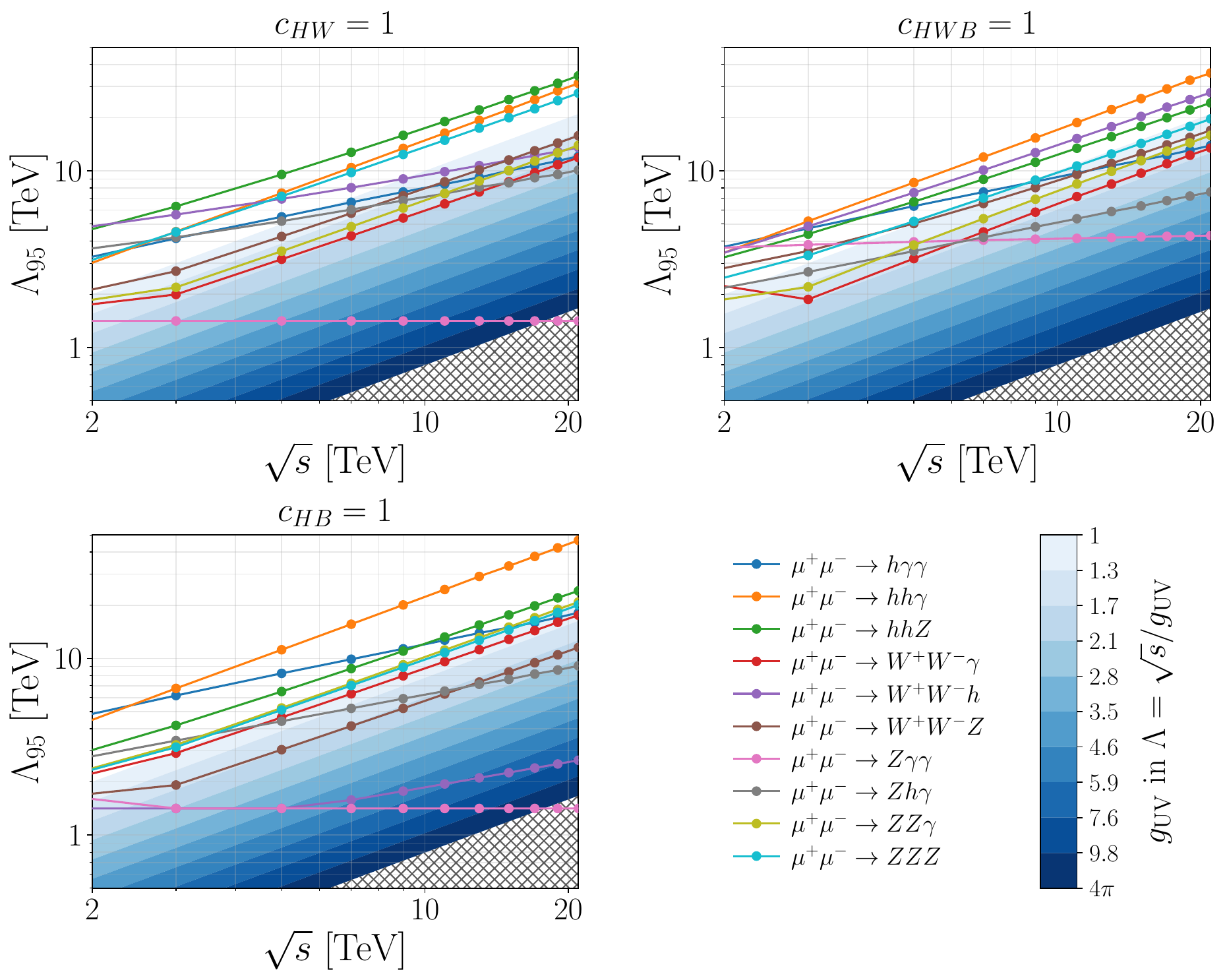}
    \caption{Same as \Cref{fig:app:twototwo-eft} for all triboson processes analysed in this work.
    \label{fig:app:twotothree-eft}}
\end{figure}

In \Cref{fig:app:twototwo-eft} and \Cref{fig:app:twotothree-eft} we show the 95\% CL lower bounds on the EFT scale $\Lambda$, as a function of $\sqrt{s}$, for all diboson and triboson processes. As discussed in the main text, the slope of the lines in this figure determines the possibility to use a process to set a constraint on the EFT at arbitrarily large energies. The blue reference lines of fixed $g_{\rm UV} = \sqrt{s}/\Lambda$ 
%of order one and $4\pi$ allows to find a 
give the maximal energy up to which the EFT assumptions may be tenable in ultraviolet completions with typical coupling $g_{\rm UV}$. Perturbative unitarity bounds of the EFT correspond to $g_{\rm UV} = 4\pi$ ($g_{\rm UV} = 1$) for tree-level (loop-level) UV completions.

%%%%%%%%%%%%%%%%%%%%%%%%%%%%%%%%%%%%%%%%%%%%%%%%%%%%%%%%%%%%%%%%%%%%%%%%%%%%%%%%%%%%%%%%%%%%%%%%%%%%
%%%%%%%%%%%%%%%%%%%%%%%%%%%%%%%%%%%%%%%%%%%%%%%%%%%%%%%%%%%%%%%%%%%%%%%%%%%%%%%%%%%%%%%%%%%%%%%%%%%%

%%%%%%%%%%%%%%%%%%%%%%%%%%%%%%%%%%%%%%%%%%%%%%%%%%%%%%%%%%%%%%%%%%%%%%%%%%%%%%%%%%%%%%%%%%%%%%%%%%%%
%%%%%%%%%%%%%%%%%%%%%%%%%%%%%%%%%%%%%%%%%%%%%%%%%%%%%%%%%%%%%%%%%%%%%%%%%%%%%%%%%%%%%%%%%%%%%%%%%%%%

%%%%%%%%%%%%%%%%%%%%%%%%%%%%%%%%%%%%%%%%%%%%%%%%%%%%%%%%%%%%%%%%%%%%%%%%%%%%%%%%%%%%%%%%%%%%%%%%%%%%

%%%%%%%%%%%%%%%%%%%%%%%%%%%%%%%%%%%%%%%%%%%%%%%%%%%%%%%%%%%%%%%%%%%%%%%%%%%%%%%%%%%%%%%%%%%%%%%%%%%%

\clearpage

%%%%%%%%%%%%%%%%%%%%%%%%%%%%%%%%%%%%%%%%%%%%%%%%%%%%%%%%%%%%%%%%%%%%%%%%%%%%%%%%%%%%%%%%%%%%%%%%%%%%
\bibliographystyle{JHEP}
\bibliography{bibfile.bib}
%%%%%%%%%%%%%%%%%%%%%%%%%%%%%%%%%%%%%%%%%%%%%%%%%%%%%%%%%%%%%%%%%%%%%%%%%%%%%%%%%%%%%%%%%%%%%%%%%%%%

\end{document}